\documentclass[structabstract]{aa}
\usepackage{txfonts}
\usepackage{natbib}
\bibpunct{(}{)}{;}{a}{}{,} 
\usepackage{graphicx}
\usepackage{hyperref}
\hypersetup{draft}
\inputencoding{utf8}
\usepackage{multirow}
\usepackage{color}

\begin{document}

\title{Effects of the selection function on metallicity trends in spectroscopic surveys of the Milky Way}
\author{G.~Nandakumar\inst{1}
\and M.~Schultheis \inst{1}
\and   M.~Hayden \inst{1}
\and  A.~Rojas-Arriagada \inst{1,3,4}
\and  G.~Kordopatis \inst{1}
\and M.~Haywood \inst{2}
}

 \institute{ Laboratoire Lagrange, Universit\'e C\^ote d'Azur, Observatoire de la C\^ote d'Azur, CNRS, Blvd de l'Observatoire, F-06304 Nice, France
 e-mail: govind.nandakumar@oca.eu
 \and
GEPI, Observatoire de Paris, PSL Research University, CNRS, Univ Paris Diderot, Sorbonne Paris Cit\'e, Place Jules Janssen, 92195 Meudon, France
 \and
 Instituto de Astrof\'{i}sica, Facultad de F\'{i}sica, Pontificia Universidad Cat\'olica de Chile, Av. Vicu\~na Mackenna 4860, Santiago, Chile \label{instA1}
\and
Millennium Institute of Astrophysics, Av. Vicu\~{n}a Mackenna 4860, 782-0436 Macul, Santiago, Chile \label{instA2}
}
\abstract { Large spectroscopic Galactic surveys imply a selection function in the way they performed their target selection. }
{ We investigate here the effect of
the selection function on the metallicity distribution function (MDF)  and on the vertical metallicity gradient by studying similar  lines of sight using four different spectroscopic surveys
(APOGEE, LAMOST, RAVE, and Gaia-ESO), which have different targeting strategies and therefore different selection functions. }
{ We use common fields between the spectroscopic surveys of APOGEE, LAMOST, RAVE  (ALR) and APOGEE, RAVE, Gaia-ESO (AGR) and use two stellar population synthesis models, GALAXIA and TRILEGAL, to create mock fields for each survey. We apply the selection function in the form of colour and magnitude cuts of the respective survey to the mock fields to replicate the observed source sample. We make a basic comparison between the models to check which  best reproduces the  observed sample distribution. We carry out a quantitative comparison between  the synthetic MDF from the mock catalogues  using both models to understand the effect of the selection function on the MDF and on the vertical metallicity gradient. }
{Using both models, we find a negligible effect of the selection function on the MDF for APOGEE, LAMOST, and RAVE. We find a negligible selection function effect on the vertical metallicity gradients as well, though GALAXIA and TRILEGAL have steeper and shallower slopes, respectively, than the observed gradient. After applying correction terms on the metallicities of RAVE and LAMOST with respect to our reference APOGEE sample, our observed vertical metallicity gradients between the four surveys are consistent within 1-$\sigma$. We also find consistent gradient for the combined sample of all surveys in ALR and AGR. We estimated a mean vertical metallicity gradient of -0.241$\pm$0.028\,dex kpc$^{-1}$. There is  a significant scatter in the estimated gradients in the literature, but our estimates are within their ranges. }
{We have shown that there is a negligible selection function effect on the  MDF and  the vertical metallicity gradients for APOGEE, RAVE, and LAMOST using two stellar population synthesis models. Therefore, it is indeed possible to combine common fields of different surveys in studies using MDF and metallicity gradients provided their metallicities are brought to the same scale. }


\titlerunning{Selection function effect on metallicity trends in the spectroscopic surveys}
\authorrunning{Nandakumar et al.}

\maketitle

\section{Introduction}

The Milky Way is the primary laboratory where we can obtain the detailed chemical, kinematic, and spatial distribution of individual stars that make up the different components (thin and thick disc, bulge, halo) of a typical spiral galaxy. Stellar atmospheres retain the composition of chemical elements present in the interstellar medium at the time and place of their formation. Thus tracing chemical abundances of individual stars combined with their kinematic properties and current phase-space location helps  us to model and test various formation and evolution scenarios of the Milky Way components to which they belong.

Metallicity is a  crucial ingredient used to decipher the Milky Way's chemical history (e.g. \citealt{2002ARA&A..40..487F,2012ARA&A..50..251I}). The mean metallicity of stellar populations is found to vary with  galactocentric radius and height from the Galactic mid-plane \citep{2015ApJ...808..132H,2012ApJ...761..160S}. The radial and vertical metallicity gradients observed in the Milky Way \citep{1982AJ.....87.1679H,2011A&A...535A.107K,2012ApJ...746..149C,2014AJ....147..116H,2014A&A...572A..33M,2016AN....337..922C,2016A&A...591A..37J} are strong signatures of formation and evolution of the different   substructures of the Milky way. Different disc formation scenarios explaining the observed abundance distribution are proposed: a radial gradient may result from the inside-out disc formation and be partially erased by radial mixing, while vertical gradients can be generated via disc heating by spiral arm interaction and mergers (see \citealt{1993ApJ...403...74Q,2001ApJ...554.1044C,2009MNRAS.396..203S,2009MNRAS.399.1145S,2013A&ARv..21...61R,2014A&A...572A..33M},  etc.). 

Tracers like open clusters, HII regions, Cepheid variables, FGK dwarfs, planetary nebulae, and red giant field stars have been used to determine the radial/vertical gradients \citetext{\citealp[etc.]{2003AJ....125.1397C,2004A&A...423..199C,2006AJ....131.2256Y,2006AJ....132..902L,2010IAUS..265..317M,2010ApJ...714.1096S,2011ApJ...738...27B,2011A&A...535A.107K,2012AJ....144..185C,2013ApJ...777L...1F,2013A&A...550A.125G,2014A&A...565A..89B,2014AJ....147..116H}}. Generally, these studies yield negative slopes for both radial and vertical gradients. However, there is a significant scatter among the estimated gradients. The radial gradient is found to vary from -0.028\,dex kpc$^{-1}$ (\citealt{2014A&A...572A..33M} for thin disc stars) to -0.17\,dex kpc$^{-1}$ (\citealt{2008A&A...488..943S} using open clusters) in the inner disc (Galactocentric radius, R<11 kpc) close to the Galactic mid-plane. The radial gradient is found to get shallower and become positve as we move away from the plane (\citealt{2013A&A...559A..59B,2014A&A...568A..71B,2014AJ....147..116H}). Similarly, a large dispersion in the estimated vertical metallicity gradient is found over the years, ranging from -0.112\,dex kpc$^{-1}$ (\citealt{2014A&A...568A..71B}) to -0.31\,dex kpc$^{-1}$ (\citealt{2008A&A...480...91S,2014AJ....147..116H}) for stars in the solar neighbourhood (7<R<9 kpc and $|$Z$|$<2 kpc). This large uncertainty of the observed metallicity gradient  makes it difficult to constrain  chemo-dynamical evolution models of the Milky Way (e.g.  \citealt{2000A&A...362..921H,2001ApJ...554.1044C,2015A&A...578A..87S}). Clearly there
is a pressing need on the observational side to reduce the uncertainty of this fundamental parameter for these models.

 During the last decade, the number of low, medium, and high resolution  spectroscopic surveys of stellar populations in our Galaxy have increased drastically \citep{2016ASPC..507...13W}. There are several multi-object spectroscopic surveys that have been completed or are underway, such as   the RAdial Velocity Experiment (RAVE; \citealt{2006AJ....132.1645S}), the LAMOST Experiment for Galactic Understanding and Exploration (LEGUE; \citealt{2012RAA....12..735D}), the Apache Point Observatory Galactic Evolution Experiment (APOGEE; \citealt{majewski2015}), the Gaia-ESO survey (GES; \citealt{2012Msngr.147...25G}), and the Sloan Extension for Galactic Understanding and Exploration (SEGUE; \citealt{2009AJ....137.4377Y}). They differ in spectral resolution, wavelength coverage, and in  their selected targets (giant stars, dwarf stars, clusters, etc.) based on their science goals. These unique target selection schemes can lead to biases in which stellar populations are observed, and affect measurements of the observed properties of the Milky Way; these targeting biases are known as the selection function. The selection function is defined as the fraction of objects in a certain colour and magnitude range successfully observed spectroscopically compared to the underlying stellar populations, and determines how representative the observed sample is compared to the full existing stellar population of the Milky Way.

 The target selection schemes limit the coverage of parameter space of $T_{\rm eff}$, log $g$, and $\rm [Fe/H]$ that  could potentially lead to biases while carrying out analyses that measure the gradients and metallicity distributions of certain stellar populations. \cite{2012ApJ...746..149C} and \cite{2012ApJ...761..160S} used different weighting schemes to correct for the metallicity bias introduced by the target selection in their sample of SEGUE  main-sequence turn-off stars, and  G and K dwarf stars, respectively. Meanwhile, \cite{2012ApJ...753..148B} determined a plate-dependent selection function for a  G-dwarf sample along $\sim$150 lines of sight in SEGUE using the dereddened colour–magnitude boxes. The selection effects in the APOGEE red clump (RC) sample is discussed in \cite{2014ApJ...790..127B} and \cite{2014ApJ...796...38N} using the much simpler and well-defined target selection algorithm of APOGEE \citep{zasowski2013}. \cite{2014ApJ...793...51S} constrained the kinematic parameters of the Milky Way disc using stars from RAVE and the Geneva–Copenhagen Survey (GCS; \citealt{2004A&A...418..989N}) using kinematic analytic models. The RAVE selection function was taken into account while modelling using GALAXIA \citep{2011ApJ...730....3S}. On comparing the temperature and colour distributions of RAVE stars with that predicted using GALAXIA, they found a reasonably good match except for J-K colours for stars in the low latitude fields. An extinction correction was performed to correct this. \cite{2017MNRAS.468.3368W} described the RAVE selection function in detail and studied the selection function effect on the RAVE metallicity and velocity parameters. For this, they created a mock-RAVE catalogue  using the GALAXIA stellar population synthesis model. They found that RAVE stars do not show any selection effects in  terms of kinematics and metallicities using the selection cuts in magnitude and colour of RAVE. \cite{2016AN....337..926A} created a mock sample of more than 600 solar-like oscillating red giant stars observed by both CoRoT and APOGEE based on their selection functions.  They found some small systematic biases of $\rm \pm 0.02\,dex$ in the radial gradient, most notably in the age bin 2--4\,Gyr.
Recently there have been many more attempts to provide a detailed description of the selection function for other major spectroscopic surveys (e.g. GES: \citealt{2016MNRAS.460.1131S}; LAMOST: \citealt{2015MNRAS.448..855Y}, \citealt{2012RAA....12..755C}).

In this paper, we study the effect of the selection function on the metallicity gradient and the metallicity distribution function (MDF) for four different spectroscopic surveys of the Milky Way tracing different stellar populations: APOGEE, LAMOST, RAVE, and GES. We  chose similar lines of sight considering common fields for three surveys at a time: APOGEE-LAMOST-RAVE (ALR) and APOGEE-GES-RAVE (AGR). We tried to emulate the selection function using the colour and magnitude cuts as they are defined for the respective surveys. We used two stellar population synthesis models to compare the distribution of the synthesized model sources with the original input catalogue in the respective colour-magnitude diagrams of each survey. The effects of the selection function are studied in detail by applying the selection function to the two models and comparing the MDF of the selected sources with that of the underlying sample.

The paper is structured as follows. In Section~\ref{specsurv} we describe the four spectroscopic surveys and their respective target selection schemes. Section~\ref{SBS} presents the comparison of stellar parameters between the surveys. The determination of common fields between the surveys, sample selection, and their spectro-photometric distance calculation are presented in Section~\ref{CFDD}. Section~\ref{MFSPS} describes in detail the stellar population synthesis models that we use to create mock fields and understand the selection function effect in MDF. Section~\ref{VMG} describes the determination of the vertical metallicity gradient, makes a comparison with the literature values, and discusses the influence of the selection function. The final conclusions of our study are given in  Section~\ref{Cncls}.

\section{Spectroscopic surveys} \label{specsurv}

We use the latest available data release of four large-scale spectroscopic surveys: APOGEE (DR13), RAVE (DR5), GES (DR4), and LAMOST (DR2). In this section, we describe the details of each survey. Each survey has an observing strategy and a specific target selection method  designed on the basis of the respective science goals. Since most of the target selection schemes make use of simple colour and magnitude cuts, the chosen photometric input catalogue(s) of the sources play an important role in selecting the stars to be observed for each survey. In this section, we describe the chosen input catalogues and the target selection schemes (colour and magnitude cuts) for each survey in detail.

We show the colour magnitude diagrams (CMDs) with the selection box (based on the colour and magnitude cuts) that defines the selection fraction (N$_{observed}$/N$_{photometric sample}$) for the respective fields of each survey. We further bin the selection box into smaller boxes, where the observed sources are located,  of 0.3 mag in magnitude and 0.05 mag in colour, which we call `masks'. These masks are used to create the mock fields from the stellar population synthesis models as described in detail in Section~\ref{models}.

\subsection{APOGEE} \label{APOGEE}

The Apache Point Observatory Galactic Evolution Experiment (APOGEE; \citealt{majewski2015}) is one of the four programs in the Sloan Digital Sky Survey III (SDSS-III; \citealt{2011AJ....142...72E}), which performed a three-year survey of our Galaxy  using the Sloan 2.5 m Telescope at the Apache Point Observatory (APO). APOGEE observes in the near-infrared $H$-band ($1.5\,\mu{\rm m}-1.7\,\mu{\rm m}$) at high spectral resolution ($R \sim 22 500$) and high signal-to-noise ratios, \textit{S$\rm/$N} ($>$100). Each plate has a field of view (FOV) ranging from 1-3\,\degr;  the number of visits per field varies   from 1 to $\sim$24 depending on the type and location of the field. The APOGEE Stellar Parameters and Chemical Abundances Pipeline (ASPCAP; \citealt{2016AJ....151..144G}) is used to determine the stellar parameters and chemical abundances of up to 15 elements based on a $\chi^{2}$ minimization between observed and synthetic model spectra.

We use the DR13 catalogue which has 164 558 sources \citep{2016arXiv160802013S}. We select only main survey targets, removing calibration, telluric, and ancillary targets, for a total sample of 109 376 stars.  APOGEE provides also  `calibrated' stellar parameters which were calibrated using a sample of well-studied field and cluster stars, including a large number of stars with asteroseismic stellar parameters from NASA’s Kepler mission. Using calibrated parameters implies a limit in $\rm log g < 3.8$. Nearly 16$\%$ of our sample lacks calibrated surface gravities and  nearly 4$\%$ of the sources have  no calibrated effective temperature and metallicity values. For this reason, we chose the uncalibrated ASPCAP values of fundamental stellar parameters for our study.

\subsubsection*{APOGEE selection function} \label{APG-selfun}
APOGEE has a well-defined input catalogue and colour selection scheme, as described in \cite{zasowski2013}. The Two Micron All Sky Survey (2MASS) Point Source Catalog \citep{2006AJ....131.1163S} is used as the base catalogue, and the targets are chosen based on their H-band magnitude and a colour limit to the dereddened (J-K$_{S}$)$_{0}$ colour \citep{zasowski2013}. The extinction corrections are derived using the Rayleigh Jeans Colour Excess (RJCE) method \citep{2011ApJ...739...25M}, which calculates the reddening values combining the 2MASS photometry with mid-IR data (Spitzer-IRAC Galactic Legacy Infrared Mid-Plane Survey \citep{{2003PASP..115..953B},{2009PASP..121..213C}} and Wide-field Infrared Survey Explorer \citep{2010AJ....140.1868W}), as 

\begin{equation}
A(K_{S}) = 0.918 \times (H - [4.5 \mu] - (H - [4.5 \mu]_{0})
\end{equation}
\begin{equation}
E(J-K_{S}) = 1.5 \times A(K_{S})
\end{equation}

A colour cut at (J-K$_{S}$)$_{0}$ $\geq$ 0.5 mag was used to include stars cool enough  that the stellar parameters and abundances can be reliably derived by ASPCAP \citep{2016AJ....151..144G}, and to lower the fraction of nearby dwarf star `contaminants' in the sample. For the halo fields ($|$b$|$>16\,\degr), the limit is extended to a bluer colour cut of  0.3 mag. In addition, for some dwarf-dominated halo fields, Washington+DDO51 photometry was used to choose more giants stars than dwarfs. The bit 7 of the APOGEE$\_$TARGET1 flag is set for sources that fulfil the Washington+DDO51 photometric giant star criteria (see \citealt{zasowski2013}). The general H-magnitude limit is taken to be 7$\leq$H$\leq$13.8, though the upper limit varies depending on the field and the plate design.

\begin{figure}[!htbp]
        \centering
                {\includegraphics[width=0.49\textwidth,angle=0]                                                         {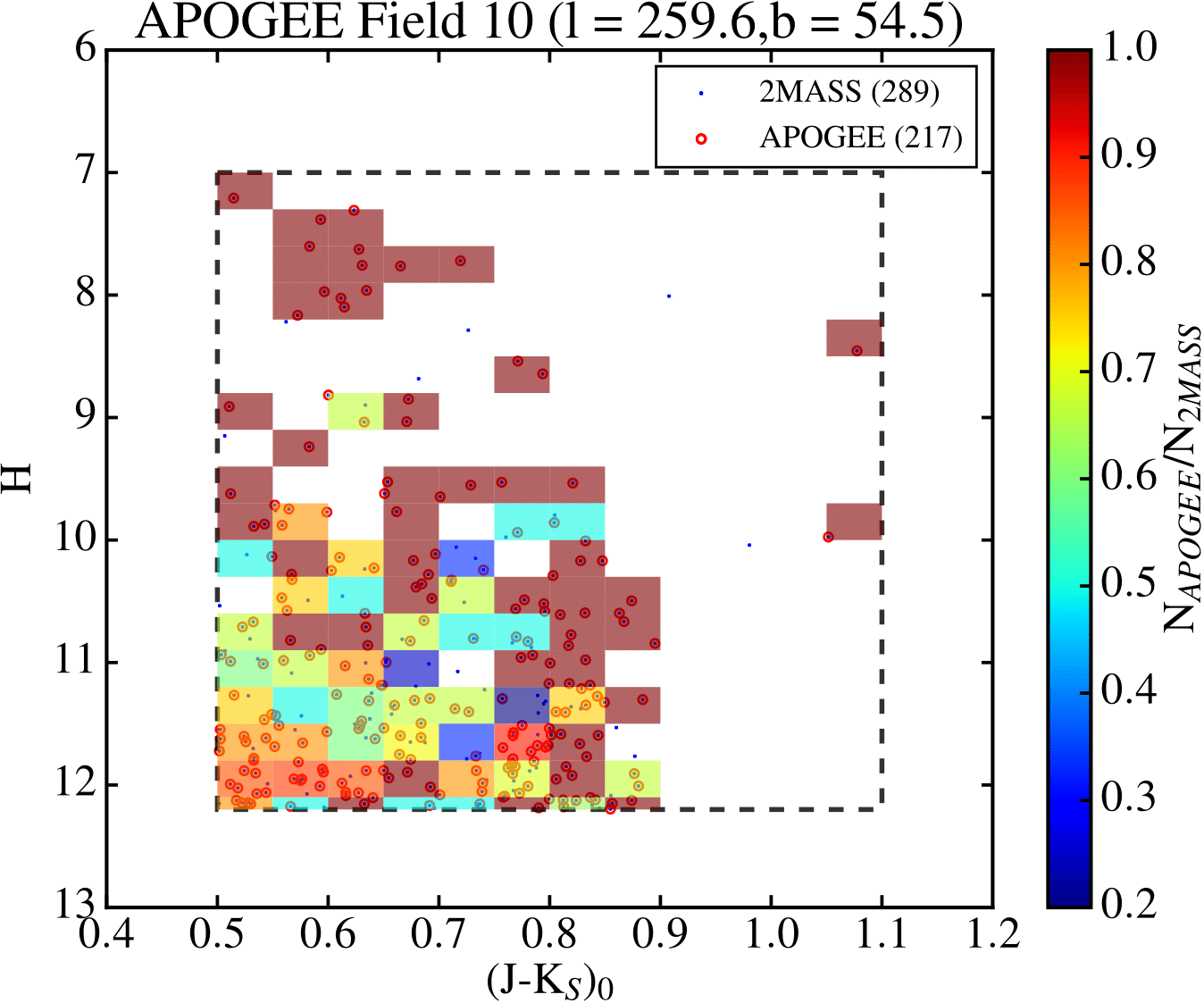}}
                \caption{ (J-K$_{S}$)$_{0}$ vs H showing the selection function for one of the fields located towards l$\sim$259.6\,\degr, b$\sim$54.5\,\degr. The bins are colour-scaled based on the N$_{APOGEE}$/N$_{2MASS}$ with sizes of 0.05 mag in (J-K$_{S}$)$_{0}$ colour and 0.3 mag in H. The  dashed box  shows the overall colour and magnitude cuts used for APOGEE.}
\label{APG_sel}
\end{figure}

Using the input 2MASS sample and their respective A(K$_{S}$) values for our fields of interest, we estimated the fraction of the observed stars with respect to 2MASS stars in small rectangular bins in the CMD. Figure~\ref{APG_sel} shows a typical example where we see that the selection function shows variations along the CMD. For the full field the fraction of observed stars to the 2MASS sample is $\sim$0.75. However, with the rectangular binning in the CMD, we see a lower fraction for fainter stars (H $>$ 11) and for bluer stars (J-K$_{S}$)$_{0}$ $<$ 0.7.

\subsection{RAVE} \label{RAVE}

The RAdial Velocity Experiment (RAVE) \citep{2006AJ....132.1645S} is a  multi-fibre spectroscopic survey that covers the entire southern celestial hemisphere except at low $|$\textit{b}$|$ and $|$\textit{l}$|$. The observations were carried out at the Anglo-Australian Observatory (AAO) in Siding Spring, Australia, using the 1.2 m UK Schmidt telescope. A 6 degree field multi-object  spectrograph was used to obtain the spectra in the infrared CaII-triplet region (8410\, \AA $<$ $\rm \lambda$ $<$ 8795\, \AA) with a spectral resolution of R$\sim$7500. The stellar atmospheric parameters were estimated using the pipeline designed for the RAVE spectra \citep{2011A&A...535A.106K,2013AJ....146..134K} making use of the MATrix Inversion for Spectral SynthEsis algorithm (MATISSE, \citealt{2006MNRAS.370..141R}) and the DEcision tree alGorithm for AStrophysics (DEGAS).

We used the DR5 catalogue of RAVE, which has more than $\sim$ 520 000 sources \citep{2017AJ....153...75K}. There are repeated observations ($\sim$15$\%$) with the same RAVEID, DENIS$\_$ID, and 2MASS$\_$ID, but which differ in the stellar parameters. In these cases, we chose the sources with the highest S/N. For our study, we used the calibrated fundamental stellar parameters which are labelled with the suffix `$\_$N$\_$K' (e.g. TEFF$\_$N$\_$K, LOGG$\_$N$\_$K, etc.; \citealt{2017AJ....153...75K}).

\subsubsection*{RAVE selection function}
For RAVE, the selection function is defined based on the I magnitude and (J-K$_{S}$) colour cut. The initial target selection is based on the apparent I-band magnitude, for 9<I<12, but the input sample is not obtained from a single catalogue. For the regions towards the Galactic disc and bulge (Galactic latitude $|$b$|$ $<$ 25\,\degr ), a colour criterion J-K$_{S}$ $\geq$ 0.5 is imposed to select cool giant stars \citep{2013AJ....146..134K}. The original input catalogue for earlier data releases of RAVE was constructed by deriving I magnitudes from the Tycho-2 catalogue \citep{2000A&A...355L..27H}, photographic I magnitudes from the SuperCOSMOS Sky Survey \citep{2001MNRAS.326.1315H}, and later using the Gunn I-band photometry from the DENIS catalogue \citep{1997Msngr..87...27E}. The latest DR4 release includes observations drawn from a new input catalogue based on DENIS DR3 \citep{2005yCat.2263....0D} cross-matched with the 2MASS point source catalogue. We adopted the same strategy as \cite{2017MNRAS.468.3368W} by choosing 2MASS as the input catalogue, and calculated an approximate I$_{2MASS}$ magnitude via the following formula:

\begin{equation}
I_{2MASS} - J = (J - K_{S}) + 0.2 \times e^{\frac{(J - K_{S}) - 1.2}{0.2}} + 0.12
\end{equation}

\begin{figure}[!htbp]
        \centering
                {\includegraphics[width=0.49\textwidth,angle=0]{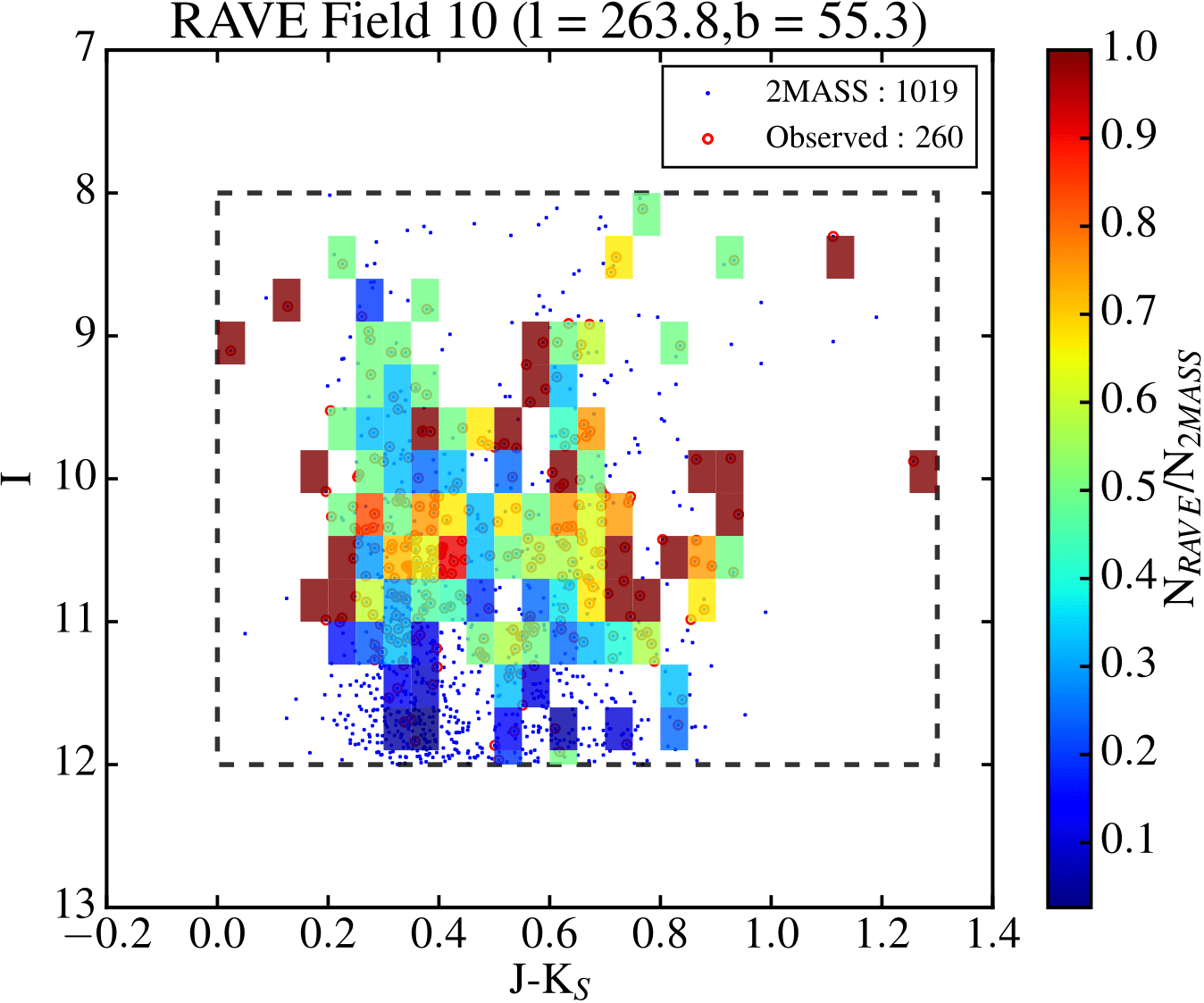}}
                \caption{ (J-K$_{S}$) vs I (CMD) showing the selection function for one of the fields located towards l$\sim$263.8\,\degr, b$\sim$55.3\,\degr. The bins are colour-scaled based on the N$_{RAVE}$/N$_{2MASS}$ with sizes of 0.05 mag in (J-K$_{S}$) colour and 0.3 mag in I. The dashed box  shows the overall colour and magnitude cuts for RAVE.}
\label{RAVE_sel}
\end{figure}   

For our fields of interest, we searched for 2MASS sources using the approximate field centres (Tables~\ref{ALR_fields} and ~\ref{AGR_fields}) with a  radius of $\rm 2.85^{o}$, and used the above-mentioned formula to calculate the I$_{2MASS}$ magnitude.  We used the same approach as in the case of APOGEE to determine the selection function for each field, i.e.  by defining selection bins in the CMD, (J-K$_{S}$) vs I$_{2MASS}$. We used sources with I$_{2MASS}$ between 8 and 12 magnitude as the I$_{2MASS}$ distribution for DR5 is shown to be broader than I$_{DENIS}$ \citep{2017MNRAS.468.3368W}. Figure~\ref{RAVE_sel} shows the CMD for one such field located towards high galactic latitude. Here again, we see a clear drop in the selection fraction to about 0.1-0.2 for the fainter magnitudes (I $>$ 11).

\subsection{GES} \label{GES} 

The Gaia-ESO survey (GES) is a public spectroscopic survey aimed at targeting $\sim$10$^{5}$ stars covering the major components of the Milky way \citep{2012Msngr.147...25G}. The observations are carried out using the Fibre Large Array Multi Element Spectrograph (FLAMES) \citep{2002Msngr.110....1P} on the Very Large Telescope array (VLT) in Cerro Paranal, Chile. This fibre facility has a FOV of 25 arcmin diameter for two different spectrographs, GIRAFFE and UVES. The stellar parameters were  derived by different nodes (using the MATISSE, SME: \citealt{1996A&AS..118..595V} and FERRE: \citealt{2006ApJ...636..804A} codes for GIRAFFE spectra, and about a dozen  different methods for UVES spectra). The final recommended GES parameters come from careful homogenization and calibration of the different results for a given star.

For our study, we chose the sources observed using GIRAFFE with two set-ups,  HR10 ($\rm \lambda $ = 5339-5619\, \AA, R$\sim$19,800) and HR21 ($\rm \lambda $ = 8484-9001\, \AA, R$\sim$16,200). We used the homogenized set of parameters from the three nodes in DR4, which are available on the public ESO webpage \footnote{http://www.eso.org/gi/}, and were  left with 29 591 sources.

\subsubsection*{GES selection function}
The GES selection function is defined in \cite{2016MNRAS.460.1131S} based on the VISTA Hemisphere Survey (VHS, \citealt{2013Msngr.154...35M}) magnitudes. We obtained the VHS catalogue for our fields of interest from the \href{http://archive.eso.org/wdb/wdb/adp/phase3$\_$main/form?phase3$\_$collection=VHS$\&$release$\_$tag=1}{ESO archive} by searching using the field centre and  a search radius of $\rm 0.2^{o}$. We used the APERMAG4 magnitude in the VHS catalogue for the corresponding J, H, and K magnitudes as it gives the closest magnitudes to those provided in the GES catalogue.\\

The target selection scheme of GES is built on stellar magnitudes and colours by defining two boxes, one blue  and one red. The blue box is used for the selection of the turn-off and main-sequence targets, while the red box is for the red clump targets \citep{2016MNRAS.460.1131S}. The colour and magnitude cuts for the blue and red boxes are given below:
\begin{itemize}
\item Blue box : 0.0 $\leq$ (J - K$_{S}$) $\leq$ 0.45 for 14.0 $\leq$ J $\leq$ 17.5
\item Red box : 0.4 $\leq$ (J - K$_{S}$) $\leq$ 0.70 for 12.5 $\leq$ J $\leq$ 15.0    
\end{itemize} 

That said, the actual target selection scheme also takes into account the extinction by shifting the boxes by 0.5$\times$E(B - V), where E(B - V) is taken as the median reddening in the field measured from the \cite{1998ApJ...500..525S} maps. Furthermore, additional targets were assigned by relaxing the red edge of the colour-cut if enough targets were not available within the colour cuts (e.g. high latitude Milky Way fields). Thus, the target selection scheme becomes 
\begin{itemize}
\item Blue box: 0.5E(B - V) + [0.0 $\leq$ (J - K$_{S}$) $\leq$ 0.45] for 14.0 $\leq$ J $\leq$ 17.5
\item Red box: 0.5E(B - V) + [0.4 $\leq$ (J - K$_{S}$) $\leq$ 0.70] for 12.5 $\leq$ J $\leq$ 15.0
\item Extra box: 0.5E(B - V) + [0.0 $\leq$ (J - K$_{S}$) $\leq$ 0.45 + $\bigtriangleup_{G}$] for J$\geq$ 14.0 and J + 3$\times$((J - K$_{S}$) - 0.35) $\leq$ 17.5
\end{itemize}

\begin{figure}[hbt!]
        \centering
                {\includegraphics[width=0.49\textwidth,angle=0]{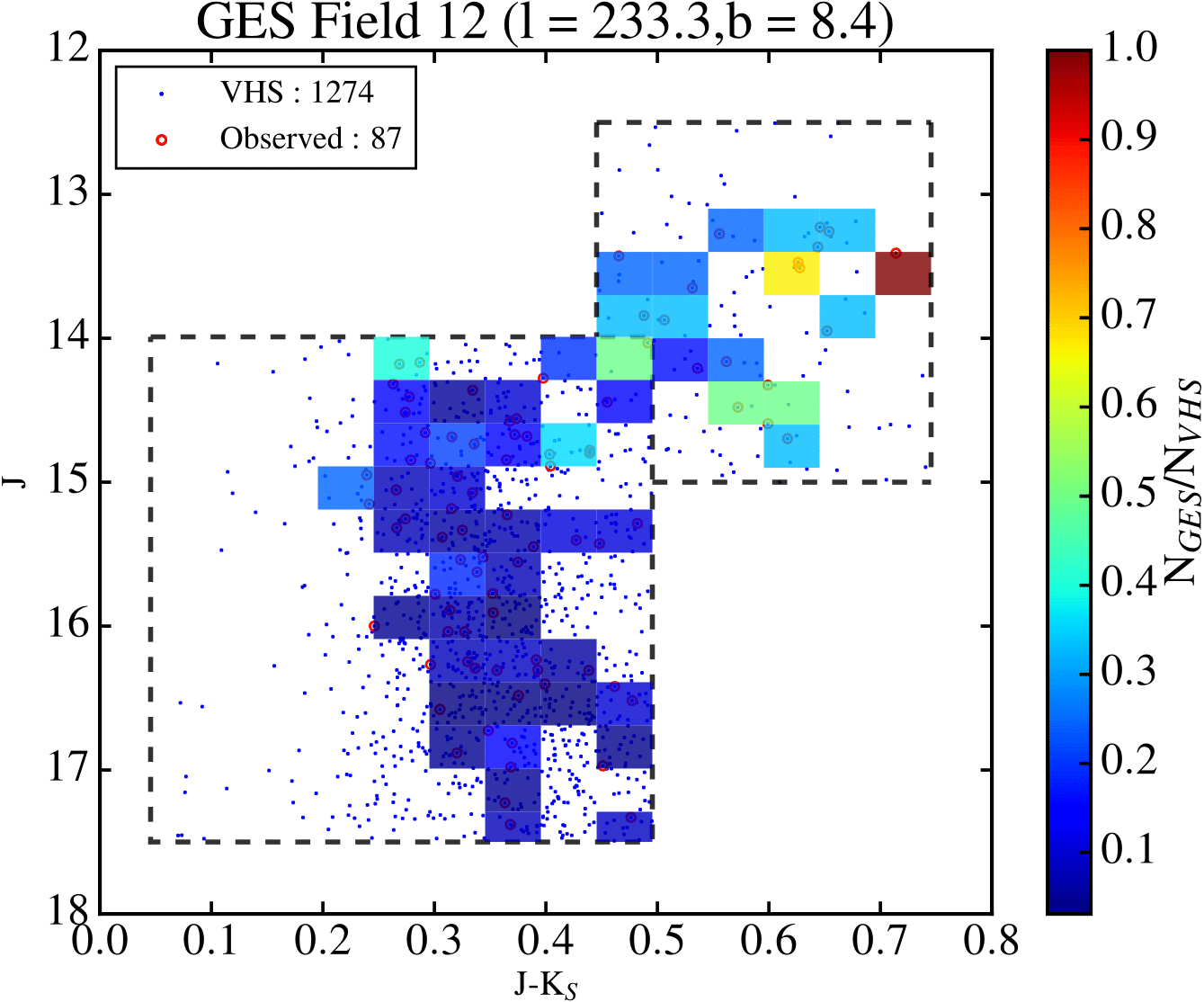}}
\caption{ (J-K$_{S}$) vs J (CM diagram) showing the selection function for  field 12 located  approximately towards l$\sim$233.3\,\degr, b$\sim$8.4\,\degr. The bins are colour-scaled based on the N$_{GES}$/N$_{VHS}$ with sizes of 0.05 mag in (J-K$_{S}$) colour and 0.3 mag in J. The dashed box  shows the overall colour and magnitude cuts for red and blue boxes used for GES.}
\label{GES_sel}
\end{figure}

where $\bigtriangleup_{G}$ is the right-edge extension of the extra box, and the values of $\bigtriangleup_{G}$ and E(B - V) are  provided in  Table 1 of \cite{2016MNRAS.460.1131S} for the required fields. Figure~\ref{GES_sel} shows the selection scheme for field 12, which is a  low latitude field  with a higher stellar density, that leads to a lower selection fraction. In this field, we  found no stellar
parameters for about 15$\%$ of the GES sources. This fraction can increase  to about 40$\%$ in other fields.

\subsection{LAMOST} \label{LAMOST} 

The Large Sky Area Multi-Object Fiber Spectroscopic Telescope (LAMOST) is another extensive ground-based spectroscopic survey of the Galaxy being carried out with the Guo Shou Jing reflecting Schmidt Telescope. It is equipped with 16 low resolution spectrographs capable of recording the spectra of up to 4000 objects simultaneously in a FOV of 5\,\degr, covering all optical wavelengths with a spectral resolution of $\sim$1800 \citep{2012RAA....12.1197C,2012RAA....12..723Z}. The survey contains  the LAMOST ExtraGAlactic Survey (LEGAS) and the LAMOST Experiment for Galactic Understanding and Exploration (LEGUE :  \citealt{2012RAA....12..735D,2012RAA....12.1021S}), which itself is composed of three separate surveys with different input catalogues and selection functions.

We use the DR2 catalogue, which has the calibrated stellar parameters for 2 207 803 sources estimated using the LAMOST Stellar Parameter Pipeline (LASP, \citealt{2011RAA....11..924W}).

\subsubsection*{LAMOST selection function}
Like RAVE, LAMOST does not make use of a single input catalogue and unlike the other three surveys, LAMOST uses \textit{ugriz} photometry for their target selection. Catalogues such as UCAC4 \citep{2013AJ....145...44Z} and Pan-STARRS 1 \citep{2012ApJ...750...99T} have been used to select targets for the observing plates in the main survey regions. To make sure that we use a homogeneous photometric input sample, we use only the SDSS photometry\footnote{2MASS was more complete in each field, but the LAMOST sources are fainter than the reliable magnitude limits ($\sim$14.3) of 2MASS pass band magnitude.} in order to define the selection function. For the respective fields, we searched for SDSS sources within a radius of $\rm 2.5^{o}$ around the field centres. We use only those fields where we have the full SDSS footprint covered. 

The targeting algorithm for LEGUE designed by \cite{2012RAA....12..755C} was not applied, due to sparse stellar sampling. This results from the limited dynamic range of magnitudes observed on a single LAMOST plate and from  the brighter r magnitude limit at the  faint end compared to the designed goal \citep{2017arXiv170107831L}. Finally, the target selection was carried out on a plate-by-plate basis with different plates covering 9 < r < 14, 14 < r < 16.8, r < 17.8, and r < 18.5.

\begin{figure}[hbt!]
        \centering
                {\includegraphics[width=0.49\textwidth,angle=0]{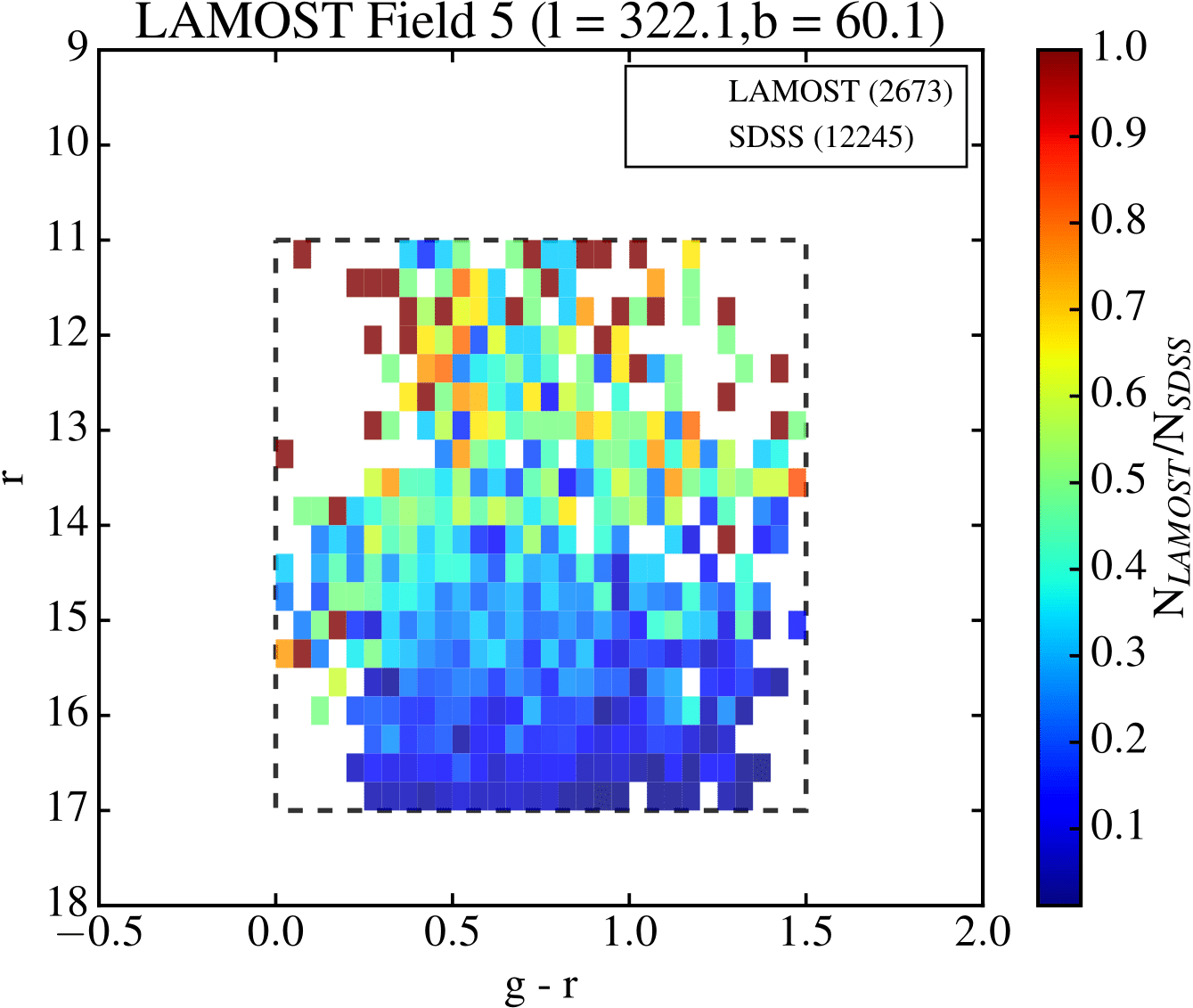}}
\caption{ (g-r) vs r (CM diagram) showing the selection function for one of the fields located towards approximate l$\sim$322.1$^{o}$, b$\sim$60.1$^{o}$. The bins are colour-scaled based on the N$_{LAMOST}$/N$_{SDSS}$ with sizes of 0.05 mag in (g - r) colour and 0.3 mag in r. The dashed box  shows the overall colour and magnitude cuts used for LAMOST.}
\label{Lam_sel}
\end{figure}

Based on the distribution of LAMOST sources of each field in the  g-r vs r CMD, we used the following colour and magnitude cuts : 0.0 < g-r < 1.5 and 11 < r < 17.8. We neglected the small fraction of very red (g-r>1.5) and blue (g-r<0.0) sources. Figure~\ref{Lam_sel} shows a typical example for a field towards l = 322.1\,\degr, b = 60.1\,\degr. Statistically, LAMOST is more prominent than the other three surveys, and this is seen in the dramatic rise in the number of masks within the colour and magnitude range. The gradual decrease in the selection fraction towards fainter magnitudes is seen here as well.

\section{Comparison of stellar parameters between the surveys} \label{SBS}

Despite the large amount of spectroscopic data available,  few studies exist (e.g. \citealt{2017A&A...600A..14S}, \citealt{2017AJ....153...75K}, \citealt{Lee2015}, \citealt{2015RAA....15.1125C}, \citealt{2013AJ....146..134K}) comparing stellar parameters and chemical abundances of the same stars. \cite{2015RAA....15.1125C} studied common stars between APOGEE DR12 \citep{2015ApJS..219...12A} and the LAMOST DR2 catalogue. They found that the LAMOST photometrically calibrated $T_{\rm eff}$ were consistent with the spectroscopic $T_{\rm eff}$ from APOGEE, but found systematic biases in log $g$ and $\rm [Fe/H]$ for common stars in the APOGEE DR12 and LAMOST DR2 catalogues. \cite{Lee2015} applied the SEGUE Stellar Parameter Pipeline (SSPP : \citealt{2008AJ....136.2070A,2008AJ....136.2022L,2008AJ....136.2050L,2011AJ....141...89S,2011AJ....141...90L}) to the spectra from LAMOST and compared the stellar parameters with the common stars in APOGEE (DR12), RAVE (DR4), and SEGUE. For the RAVE DR5 release \citep{2017AJ....153...75K} a detailed comparison of the derived stellar parameters in RAVE with that of APOGEE, GES, and LAMOST for the common stars has been obtained as a part of the external RAVE DR5 verification. 

Recently, a data-driven approach known as \textit{The Cannon} \citep{2015ApJ...808...16N} has been introduced. \textit{The Cannon} uses stellar spectra along with the derived stellar parameters from well-characterized stars (estimated with pipelines using synthetic model spectra) in higher resolution surveys as a training set    to derive stellar parameters. This method was used to derive the stellar parameters for around 450 000 giant stars in LAMOST (low spectral resolution survey) by bringing them to the scale of APOGEE (high spectral resolution), showing that two very different spectroscopic surveys can be combined together \citep{2017ApJ...836....5H}. But still there are limitations in this data-driven approach, as the accuracy of \textit{the Cannon} depends on the chosen training set. In the cases of APOGEE and LAMOST, which target different populations -- red giants stars for APOGEE vs dwarf stars for LAMOST -- only a limited training set is available. \textit{The Cannon} was also used to re-analyse the RAVE spectra and a new catalogue (RAVE-on) of stellar parameters and abundances was produced \citep{2017ApJ...840...59C}. The training set for red giant stars was made using common stars in APOGEE DR13 \citep{2016arXiv160802013S}, while the stars in common in K2/EPIC catalogue \citep{2016yCat..22240002H} made up the main-sequence and subgiant branch training set.

 For our study, we need to make sure that systematic offsets between surveys are corrected. This is accomplished by comparing the derived stellar parameters for the common sources between the surveys. We have arbitrarily chosen the APOGEE data set as a reference frame due to its high spectral resolution and S/N. We used a cross-identification radius of 2'' to identify the common sources  of the three  surveys with respect to APOGEE.

 In this section, we investigate the systematic offsets seen in $T_{\rm eff}$, log $g$, and $\rm [Fe/H]$\footnote{APOGEE, RAVE, and GES use the $\rm [M/H]$ notation for the overall content of metallic elements, rather than $\rm [Fe/H]$. However, throughout this paper, we use $\rm [Fe/H]$ as the the global metallicity, i.e. $\rm [Fe/H]$ = $\rm [M/H]$. For LAMOST, only $\rm [Fe/H]$ is provided in the catalogue.} between APOGEE and the other data sets,  RAVE, GES, and LAMOST, respectively. We found only six sources in common between GES and APOGEE, making any comparison difficult. Therefore, we carried out the comparison only between APOGEE and RAVE and between APOGEE and LAMOST. Table~\ref{sbs_tab} lists the median offsets and the standard deviation estimated for APOGEE - RAVE and APOGEE - LAMOST.
 
 \begin{figure*}[!htbp]
  \centering
        \includegraphics[width=\textwidth,angle=0]{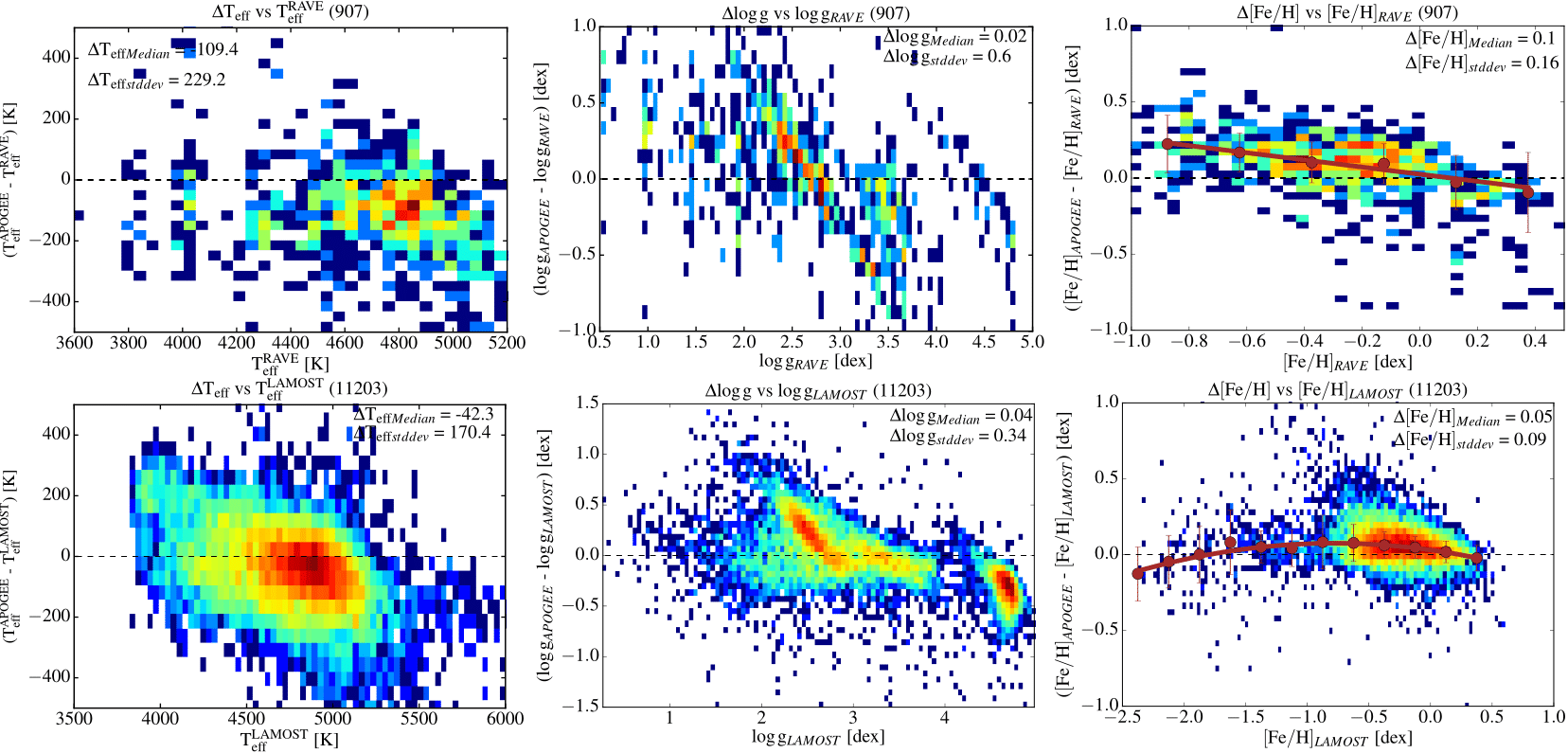}
        \caption{Comparison of common sources in APOGEE and RAVE (Top), and APOGEE and LAMOST (Bottom) for $T_{\rm {eff}}$ (left), $\log{g}$ (middle), and [Fe/H] (right). For the metallicities, the median of the difference in parameters and its dispersion is also shown as red circles with error bars in the plots. $\rm [Fe/H]$ here denote the global metallicity for APOGEE and RAVE (see footnote 2). }
  \label{APG_ALL_img}
  \end{figure*}

  \begin{table}[hbt!]
\begin{center}
\caption{ Median offset and dispersion estimated for the comparison of fundamental stellar parameters in APOGEE - RAVE and APOGEE - LAMOST. }
\label{sbs_tab}
\vspace{0.2cm}
\begin{tabular}{l c c c}
\hline
\\
\multicolumn{4}{c}{APOGEE - RAVE (907)}  \\
\hline
& & Median Offset & Dispersion\\
\hline
\multirow{ 3}{*} & $T_{\rm eff}$ (K) & -109.4 & 229.2 \\
& Log $g$ (\,dex) &   0.02 &  0.60 \\
& $\rm [Fe/H]$ (\,dex) & 0.10 & 0.16 \\

\hline
\\
\multicolumn{4}{c}{APOGEE - LAMOST (11 203)}  \\
\hline
& & Median Offset & Dispersion\\
\hline
\multirow{ 3}{*} & $T_{\rm eff}$ (K) & -42.3 & 170.4 \\
& Log $g$ (\,dex) &   -0.04 & 0.34 \\
& $\rm [Fe/H]$ (\,dex) & 0.05 & 0.09 \\

\hline

\end{tabular}
\end{center}
 \end{table}

\subsection{APOGEE-RAVE} \label{APG-RAVE}

There are 907 sources in common between APOGEE and RAVE. Figure~\ref{APG_ALL_img} (top panel) shows the comparison of the stellar parameters. We find that RAVE has systematically higher temperatures than APOGEE with a median difference of about 110\,K. The log $g$ values show a peculiar shape (Figure~\ref{APG_ALL_img}, top  middle panel), which has been already noted by \cite{2017AJ....153...75K}. They consider this behaviour to be the consequence of degeneracies in the Ca IR triplet region that affects the determination of log $g$ \citep{2011A&A...535A.106K}. Overall we see a large dispersion (0.6\,dex) for log $g$ between APOGEE and RAVE. In terms of the metallicity comparison, we note that APOGEE gives higher metallicities for metal-poor stars ($\rm [Fe/H]$ < -0.2\,dex), while much lower metallicities for metal-rich stars ($\rm [Fe/H]$ > 0.2\,dex) in comparison with RAVE. We have calculated the median offsets in bins of 0.25\,dex in RAVE  metallicities (indicated by the red points)  and did a linear fit to them, as indicated by the red  line.

\subsection{APOGEE-LAMOST} \label{APG-LAM}  

        LAMOST has 11 203 sources in common with APOGEE, which is statistically the highest number. The plots used for comparison are shown in the bottom panel of Figure~\ref{APG_ALL_img}. The median offset of $T_{\rm eff}$ between LAMOST and APOGEE is about 42\,K with a dispersion of 170\,K. A slight trend with $T_{\rm eff}$ is visible in the sense that for stars hotter than 5000\,K LAMOST predicts higher temperatures. However, the APOGEE pipeline is more adapted to getting stellar parameters for cooler stars below 5000\,K \citep{2015AJ....150..148H}. The log $g$ correlation shows distinct behaviour for the APOGEE values above and below $\sim$ 4.0\,dex. Below 4.0\,dex we see in general good agreement except around log $g\sim2.5$, the area where the red clump is dominant. The log $g$ values above 4.0\,dex are underestimated in APOGEE because of the lack of reliable calibrators for stars with high surface gravity values \citep{2015AJ....150..148H}. This can be seen in the behaviour of difference in log $g$ with log\, $g_{LAMOST}$ for log $g$ $>$ 4.0\,dex. Hence, we estimate the median offset and  dispersion for log\,$g_{APOGEE}$ $<$ 4.0\,dex. Overall, the median offset is negligible ($\sim$ -0.04\,dex), but with a large dispersion $(\sim$0.34\,dex). In the case of metallicity, the median offset between APOGEE and LAMOST for sources within -1 $<$ $\rm [Fe/H]$ $<$ 0.5 is 0.05\,dex with a dispersion of 0.14\,dex. Here again, we added correction terms for the metallicities by calculating the median offset in bins of 0.25 dex in metallicity and fitting a second-degree polynomial to them (red points and line).

\section{Common fields and distance determination}  \label{CFDD}

We have chosen to select common fields distributed along similar lines of sight to study the effect of the selection function on the observed MDF and the vertical gradient.
In total, there are only three  common fields between the four surveys, which provide a relatively small sample size, and for this reason we have chosen to restrict ourselves to the common fields between three surveys at a time: APOGEE, LAMOST, and RAVE (hereafter  ALR) and APOGEE, GES, and RAVE (AGR). We found eight common fields in the ALR case and ten in AGR. Tables~\ref{ALR_fields} and ~\ref{AGR_fields} list the common fields in each survey of ALR and AGR, respectively, with the approximate mean field centres and number of sources in each field. \\

\begin{figure}[!htbp]
        \centering
                {\includegraphics[width=0.49\textwidth,angle=0]{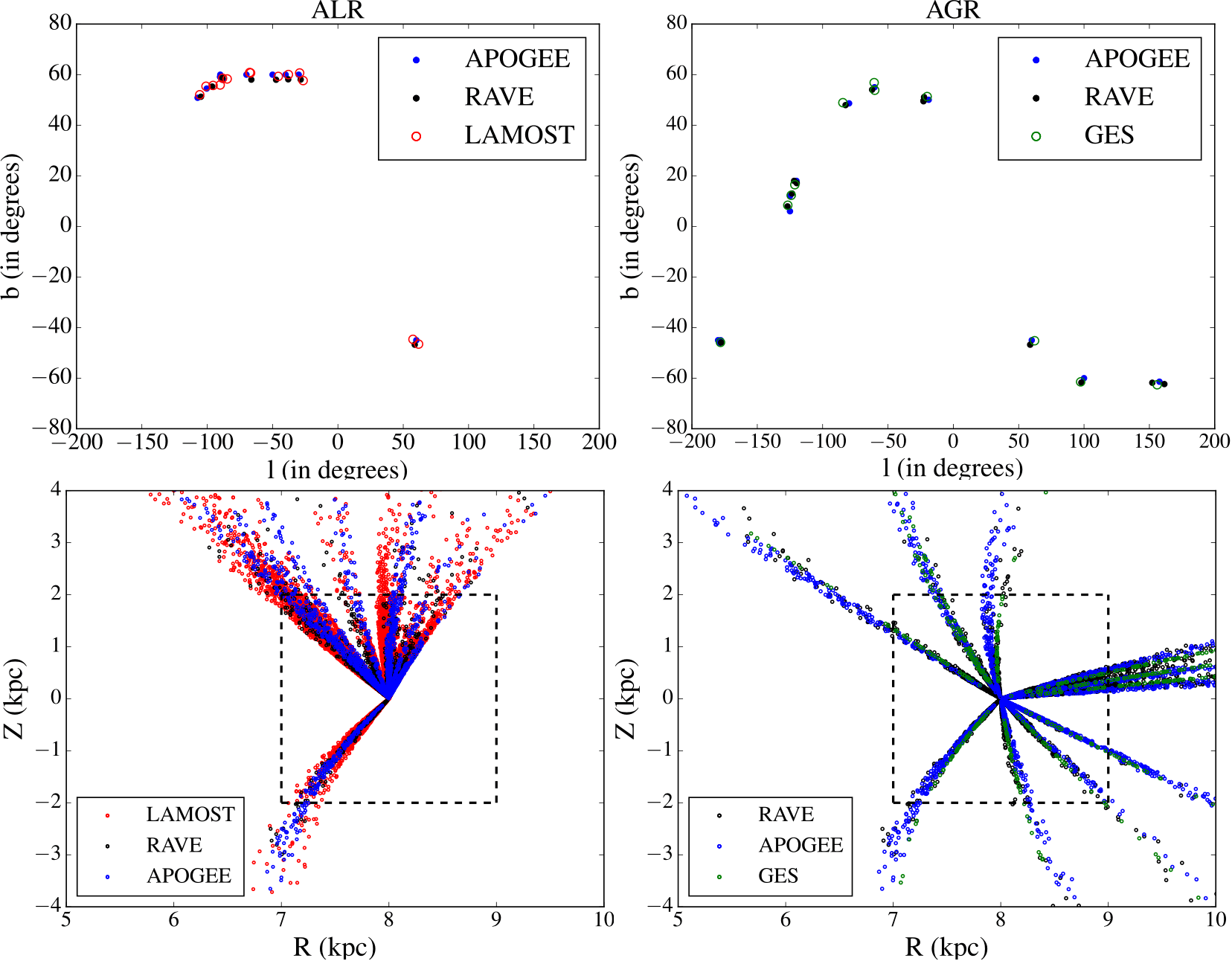}}
                \caption{Distribution of the common fields in ALR and AGR shown in the Galactic plane (top); the R-Z distribution of the sources in those fields with the dashed box indicating the 7$\leq$R$\leq$9 kpc and $|$Z$|$ $\leq$ 2 kpc range we chose to select the sources for our study  (bottom).}
\label{ALR_AGR}
\end{figure}

\begin{table*}[hbt!]
\begin{center}
\caption{Details of the ALR fields. The field numbers assigned by us, mean of field centres of each field for three surveys, along with the number of observed sources in each field are listed. The number of observed sources having their distances calculated (based on availability of derived $T_{\rm eff}$, log $g$, $\rm [Fe/H],$ and photometric magnitudes) are indicated in parentheses alongside the observed source numbers. There are overlapping fields for certain surveys, indicated by $\_$1, $\_$2, or $\_$3. }
\label{ALR_fields}
\vspace{0.2cm}
\begin{tabular}{l c c c c }
 \hline \hline 
 Field & Mean(l\,\degr,b\,\degr) & N(APOGEE) & N(RAVE)  & N(LAMOST) \\
 \hline 
 
 5 & (320.7, 59.4) & 222 (222) & 497 (495) & 2844 (2677)\\
 \\
 10$\_$1 & (262.6,55.2) & 221 (221)  & 277 (277)  & 503 (475)\\
 10$\_$2 & (259.1,55.4) & --- & --- & 1306 (1175) \\ 
 \\
 13$\_$1 & (58.7,-45.5) & 228 (228) & 566 (566) & 719 (709)\\
 13$\_$2 & (61.7,-46.5) & --- & --- & 1185 (1160) \\ 
 \\
 14 & (312.4,59.1) & 225 (225) & 250 (248)  & 1208 (777) \\ 
 \\
 15$\_$1 & (292.1,59.6) & 221 (221)  & 362 (362)  & 994 (898)\\
 15$\_$2 & (293.2,60.7) & --- & --- & 990 (846) \\ 
 \\
 20$\_$1 & (331.6,58.6) & 227 (227) & 306 (306) & 2822 (2488) \\
 20$\_$2 & (330.7,60.7)  & --- & --- & 3021 (2870)\\ 
 \\ 
 25$\_$1 & (273,58.3) & 252 (252) & 282 (282) & 2668 (2500) \\
 25$\_$2 & (269.9,57.9) & 263 (262) & ---  & 385 (360) \\
 25$\_$3 & (270.7,58.8) & 260 (260) & ---  & 1542 (1460) \\
 \\
 32 & (254,51.4) & 225 (225) & 300 (300)  & 1793 (1698) \\
 \hline
 \end{tabular}
\end{center}
 \end{table*}

\begin{table*}[hbt!]
\begin{center}
\caption{Details of the AGR fields. The columns are the same as in Table~\ref{ALR_fields}}
\label{AGR_fields}
\vspace{0.2cm}
\begin{tabular}{l c c c c }
 \hline \hline 
 Field & Mean(l\,\degr,b\,\degr) & N(APOGEE) & N(RAVE)  & N(GES) \\
 \hline 
 1$\_$1 & (339.6,50.8) & 228 (228) & 485 (485) & 103 (54)\\
 1$\_$2 & (337.1,49.7) & --- & 521 (521) & ---\\
 1$\_$3 & (337.2,49.5) & --- & 534 (534) & ---\\
 \\
 3 & (182,-45.6) & 222 (222) & 155 (155) & 91 (47)\\
 \\
 4$\_$1 & (158.4,-62.1) & 259 (259) & 389 (389) & 82 (45)\\
 4$\_$2 & (152.2,-61.8) & --- & 233 (233) & --- \\
 \\
 5$\_$1 & (239.6,17.2) & 230 (230) & 614 (614) & 104 (81) \\ 
 5$\_$2 & (238.2,18) & --- & 479 (479) & --- \\
 \\
 8 & (60.3,-45.7) & 228 (228) & 566 (566) & 103 (59)\\
 \\
 10 & (277.7,48.5) & 220 (220) & 582 (582) & 102 (51) \\
 \\ 
 12 & (233.8,7.4) & 229 (229) & 801 (801) & 104 (87) \\
 \\
 14 & (235.9,12.4) & 229 (229) & 451 (451) & 108 (88) \\
 \\
 23 & (98.4,-61.1) & 279 (279) & 320 (320) & 86 (47) \\
 \\ 
 31$\_$1 & (299.1,54.3) & 231 (231) & 348 (348) & 108 (60) \\
 31$\_$2 & (299.4,56.9) & --- & --- & 96 (54) \\
 \hline
 \end{tabular}
\end{center}
 \end{table*}

Distances for the sources in each survey are estimated by isochrone fitting as described in \cite{2017A&A...601A.140R}, which is similar to other methods in the literature (e.g. \citealt{2010A&A...522A..54Z}) using the derived stellar parameters $T_{\rm eff}$, log $g$, $\rm [M/H]$ together with the J, H, K$_{S}$ (for APOGEE, RAVE, and GES) or SDSS u,g,r,i photometry (for LAMOST). A set of PAdova and TRieste Stellar Evolution Code (PARSEC) isochrones with ages ranging from 1 to 13 Gyr in 1 Gyr step and metallicities from -2.2 dex to +0.5 dex in 0.1 dex step are chosen for this. PARSEC  is the stellar evolutionary code used to compute sets of stellar evolutionary tracks for stars of different intial masses, evolutionary phases, and metallicities. Isochrones in several photometric systems are derived from these tracks \citep{2012MNRAS.427..127B}. The distance of the observed star to the set of all model stars from the whole set of isochrones is calculated in the $T_{\rm eff}$-log $g$-$\rm [Fe/H]$ parameter space. This distance is weighted to account for the evolutionary speed and non-uniformity of model stars along the isochrone tracks. Using these weights, the most likely values of absolute magnitudes of the star in each band is calculated as the weighted mean or median of the model stars' absolute magnitudes. We also compute the line-of-sight reddening from the observed and theoretical colours. Finally, we compute the distance modulus and the line-of-sight distance in each passband from the absolute magnitudes and the estimated reddening \citep{2017A&A...601A.140R}. We use the average value of the distances from different passbands as the final line-of-sight distance of each source. Using the same approach to calculate the distances for each of the four surveys makes sure that no biases are introduced. The typical uncertainty of the distances is in the order of $\rm \sim 20\%$.
The Galactocentric distance R (kpc) and the vertical height Z (kpc) from the Galactic mid-plane for the sources are calculated by assuming the Sun to be located at R$\sim$8.0 kpc. The distribution of the fields in the Galactic plane and that of the sources in the R-Z plane are shown in Figure~\ref{ALR_AGR}. We see from the R-Z distribution in the bottom panel of Figure~\ref{ALR_AGR} that there is a wide range of Z allowing us to determine the vertical gradient (see Section~\ref{VMG}), while the range in the radial gradient is limited. Even though the range in R is broader in AGR than that in ALR, to be consistent between the two cases and to minimize the impact of any radial gradient on the vertical gradient, we restrict the samples in R from 7 to 9 kpc and $|$Z$|$ from 0 to 2 kpc for our study (see the dashed box in the bottom panel of Figure~\ref{ALR_AGR}).

\begin{figure}[!htbp]
        \centering
                {\includegraphics[width=0.49\textwidth,angle=0]{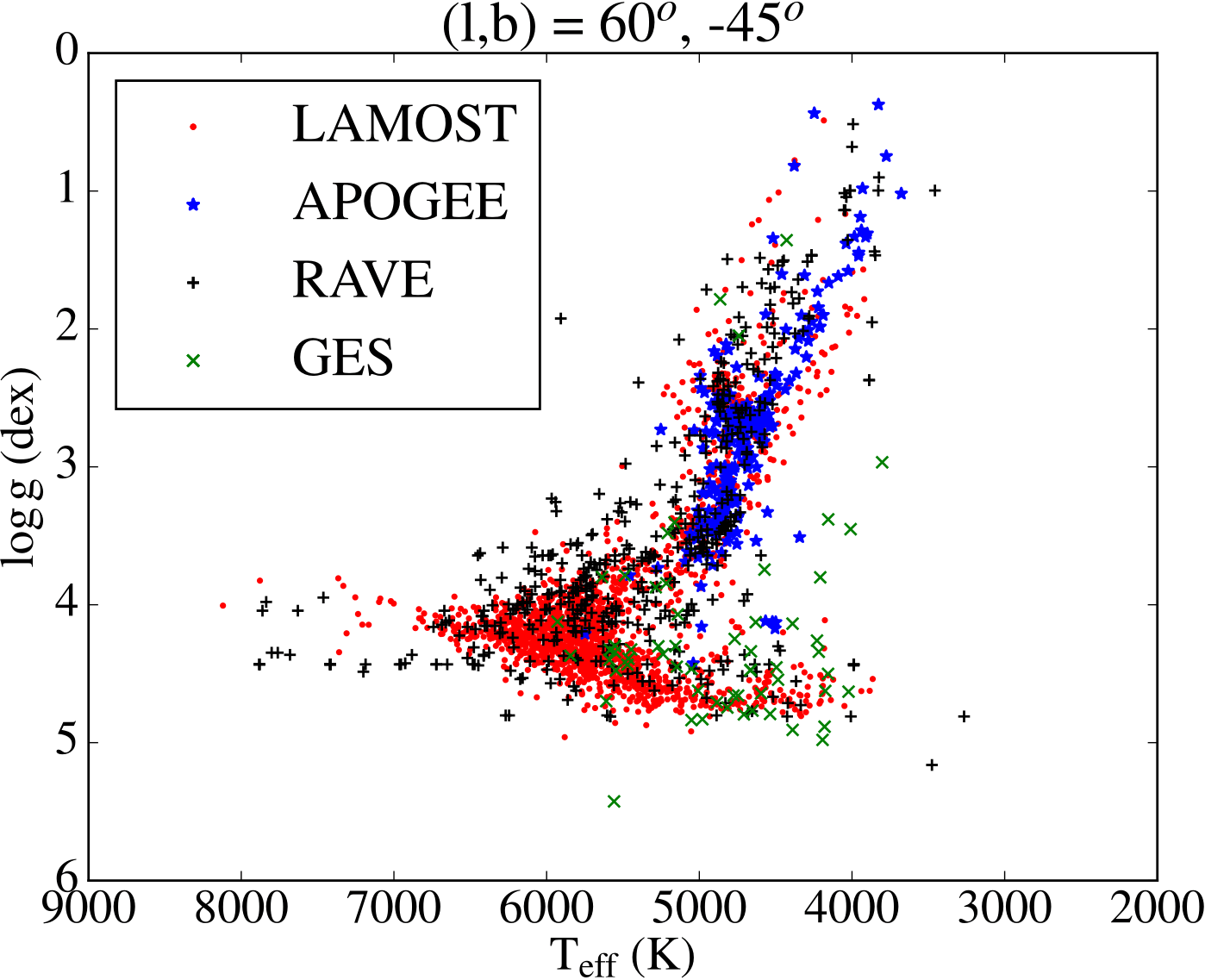}}
                \caption{ $T_{\rm eff}$ vs log $g$ diagram of sources for the four surveys in the field located towards (l,b) $\sim$ 60\,\degr, -45\,\degr. }
\label{HR}
\end{figure}

 In Figure~\ref{HR}, we show the Hertzsprung-Russell diagram for the sources in the field common to all four surveys located towards (l,b)$\sim$ 60\degr, -45\degr. The location of LAMOST and RAVE sources are mainly concentrated along the main-sequence and the turn-off stars, while APOGEE traces mainly giant stars and red clump stars. The lowest density comes from GES tracing mainly main-sequence stars. 

\section{MOCK fields using stellar population synthesis models} \label{MFSPS}

Stellar population synthesis models make use of Galaxy formation and evolution scenarios along with some physical assumptions to generate a picture of the Milky Way in different photometric systems. Their prime objective is to compare and interpret different observational data currently available and also to test the theories on which the models are based. These models use the fundamental equation of stellar statistics \citep{1986ARA&A..24..577B} to compute the star counts for the distinct components (thin and thick discs, halo, bulge) that make up the Milky Way. Fundamental stellar parameters are derived assuming each component's respective density distribution laws, inital mass function and luminosity function, star formation rate, age-metallicity relation, etc., using libraries of stellar evolutionary tracks and synthetic spectra. GALAXIA \citep{2011ApJ...730....3S} and TRILEGAL \citep{2005A&A...436..895G} are the two most commonly used population synthesis models differing in their assumed component parameters, density laws, star formation history, stellar libraries, and other dynamical constraints. We chose them to create the respective mock fields for each survey with the aim of understanding the selection function effect in the MDFs and to attempt a basic comparison of the models. We chose two stellar population synthesis models with different input physics to test the robustness of our analysis. 

GALAXIA uses a modified version of the BESAN\c CON model \citep{2003A&A...409..523R}, which is dynamically self consistent, constraining the scale height of populations (assumed to be isothermal and relaxed) by its velocity dispersion and Galactic potential. Among the four main populations, the thin disc is divided into seven age components (0 to 10 Gyr) and has a two-slope initial mass function. The thick disc has a mean metallicity of -0.78 dex simulated as a single burst of age 11 Gyr, while the halo (mean metallicity$\sim$ -1.78 dex) and bulge (mean metallicity$\sim$ 0.0\,dex) are simulated as single bursts of ages of 14 and 10 Gyr,  respectively. Instead of the Schlegel extinction maps \citep{1998ApJ...500..525S} used by GALAXIA to calculate the extinction along a given line of sight, we chose the more sophisticated model of  the 3D dust distribution provided by \cite{2003A&A...409..205D}. 

TRILEGAL does not include the dynamical consistency to constrain the scale height, but like GALAXIA it is able to deal with a full set of different photometric systems. TRILEGAL also has four galactic components with certain input parameters that can be modified, such as thin and thick discs, halo, and bulge. We assumed a thin disc with a total mass surface density of 55.4 M$_{\odot}$ pc$^{-2}$, a scale length h$_{R}$ = 2.15 kpc, and an age dependent scale height h$_{Z}$(t$_{Gyr}$) = 245(1 + t/5.5)$^{1.66}$ pc. The abundances of \cite{2000A&A...358..850R} are adopted for the thin disc and a two-step SFH with a 1.5 times enhancement in the SFR between the ages of 1 and 4 Gyr. The thick disc is assumed to have a local mass volume density of 0.008 M$_{\odot}$ pc$^{-3}$, h$_{R}$ = 3.2 kpc, and h$_{Z}$ = 0.74 kpc. The SFH is constant over an age range of 11-12 Gyr with mean metallicity $\sim$ -0.67 $\pm$ 0.1\,dex and an $\alpha$-enhancement of $\sim$ 0.3\,dex. The halo is modelled using an axisymmetric power law with a power-law index of 2.75 \citep{2010ApJ...714..663D} and local mass volume density of 4 $\times$ 10$^{-4}$ M$_{\odot}$ pc$^{-3}$. The SFH is constant over an age range of 12-13 Gyr with mean metallicity $\sim$-1.6$\pm$1.0\,dex and a corresponding $\alpha$-enhancement of $\sim$0.3\,dex. We use the \cite{2003A&A...409..205D} dust distribution to calculate the extinction in TRILEGAL as well. The model scheme and other details are described in \cite{2005A&A...436..895G}. 

We generate the mock catalogues along each line of sight using the field centre and the field radius of the respective surveys for each of the model. The 2MASS+SDSS photometric system was used for both the models. 

\subsection{Applying uncertainties and related checks on the models}

Both GALAXIA and TRILEGAL predict the stellar parameters and photometric magnitudes for each star at a given line of sight.  Each of the four surveys has intrinsic errors in the measured stellar parameters and observed photometric magnitudes, which should be simulated accordingly  in the model  in order to  make it more realistic. Since we use only the metallicity values from the models to compare the MDFs and vertical gradients, we do this only for the metallicity among the stellar parameters in the model. In order to simulate the metallicity errors, we have fitted a  fourth-degree polynomial in the $\sigma_{\rm [Fe/H]}$ vs $\rm [Fe/H]$ plane for APOGEE, LAMOST, and GES. For RAVE, we used the same metallicity error description as described in \cite{2013AJ....146..134K}. We apply a Gaussian filter to the metallicities of the mock catalogues.   

Similarly we need to apply uncertainties to the photometric 2MASS or SDSS magnitudes provided by each model. We  used an exponential function for 2MASS and a fourth-degree  polynomial for SDSS to define the relation between the mag vs $\sigma_{mag}$, which we model as a Gaussian  for each model source, as mentioned earlier.

In addition, we have simulated errors in the distance distribution in the model to verify the percentage of souces lost or gained, due to our limiting R-Z cut. The typical errors in the spectro-photometric distances are in the order of 20$\%$ (see \citealt{2015ApJ...808..132H}, \citealt{2015A&A...577A..77S}). We carried out ten trials, each time introducing a 20$\%$ error in distance calculated by the models to check whether this drastically affects the number of sources at the boundaries of the R-Z range that are thrown out or that come in. We have found that a 20$\%$ error in the distances would affect less than 5$\%$ of stars in the selected R-Z ranges. This is a relatively small change in the sample size, but to make the mock sample realistic, we kept the 20$\%$ distance uncertainty in the models.

\subsection{Comparison between GALAXIA and TRILEGAL}\label{models}

\begin{figure}[!htbp]
  \centering
        \includegraphics[width=90mm, height = 45mm,angle=0]{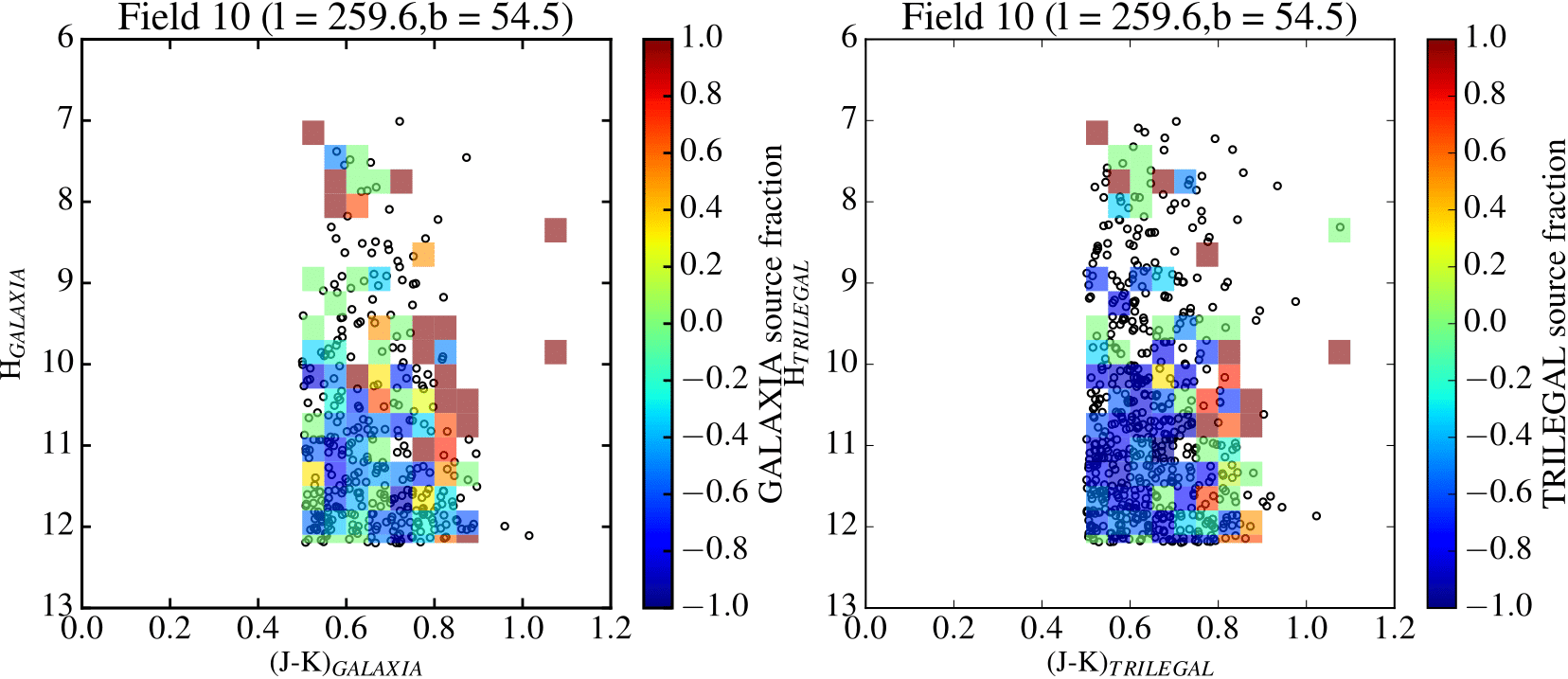}
        \caption{ CMD diagrams for the APOGEE field located towards l,b$\sim$59.6$^{o}$, 54.5$^{o}$ with GALAXIA and TRILEGAL source distribution shown in the left and right panels, respectively. The open circles in each panel represent the respective model sources. The rectangular boxes are the masks where the real observed sources are, each colour-coded with the fraction (N$_{2MASS}$ - N$_{model}$)/N$_{2MASS}$ if N$_{2MASS}$ $>$ N$_{model}$ or (N$_{2MASS}$ - N$_{model}$)/N$_{model}$ if N$_{2MASS}$ $<$ N$_{model}$. The redder colours indicate that the 2MASS sources are in equal number or greater than the number of model sources, while bluer colours denote more model sources. From the colours of the bins, there are more TRILEGAL sources than GALAXIA sources towards the faint magnitudes. }
  \label{BES_TRI}
  \end{figure}

In Section~\ref{specsurv}, we show the CMDs for each survey with their respective colour and magnitude cuts that were used to select the target sample. We carry out the same exercise for the mock fields generated using the models. The masks (in the form of small rectangular boxes within the selection box) are used to denote the bins in colour and magnitude where the sources are observed. Figure~\ref{BES_TRI} shows an example of our method for the APOGEE field located towards l$\sim$259.6$^{o}$ and b$\sim$54.5$^{o}$ for both GALAXIA and TRILEGAL. The masks are colour-coded  with the fractional percentage of model sources compared to the input photometric sample (2MASS or SDSS). We find that TRILEGAL predicts more sources at the faint end than does GALAXIA. A very similar trend is seen for the simulated RAVE, LAMOST, and GES fields.
  
\begin{figure*}[!htbp]
  \centering
        \includegraphics[width=\textwidth,angle=0]{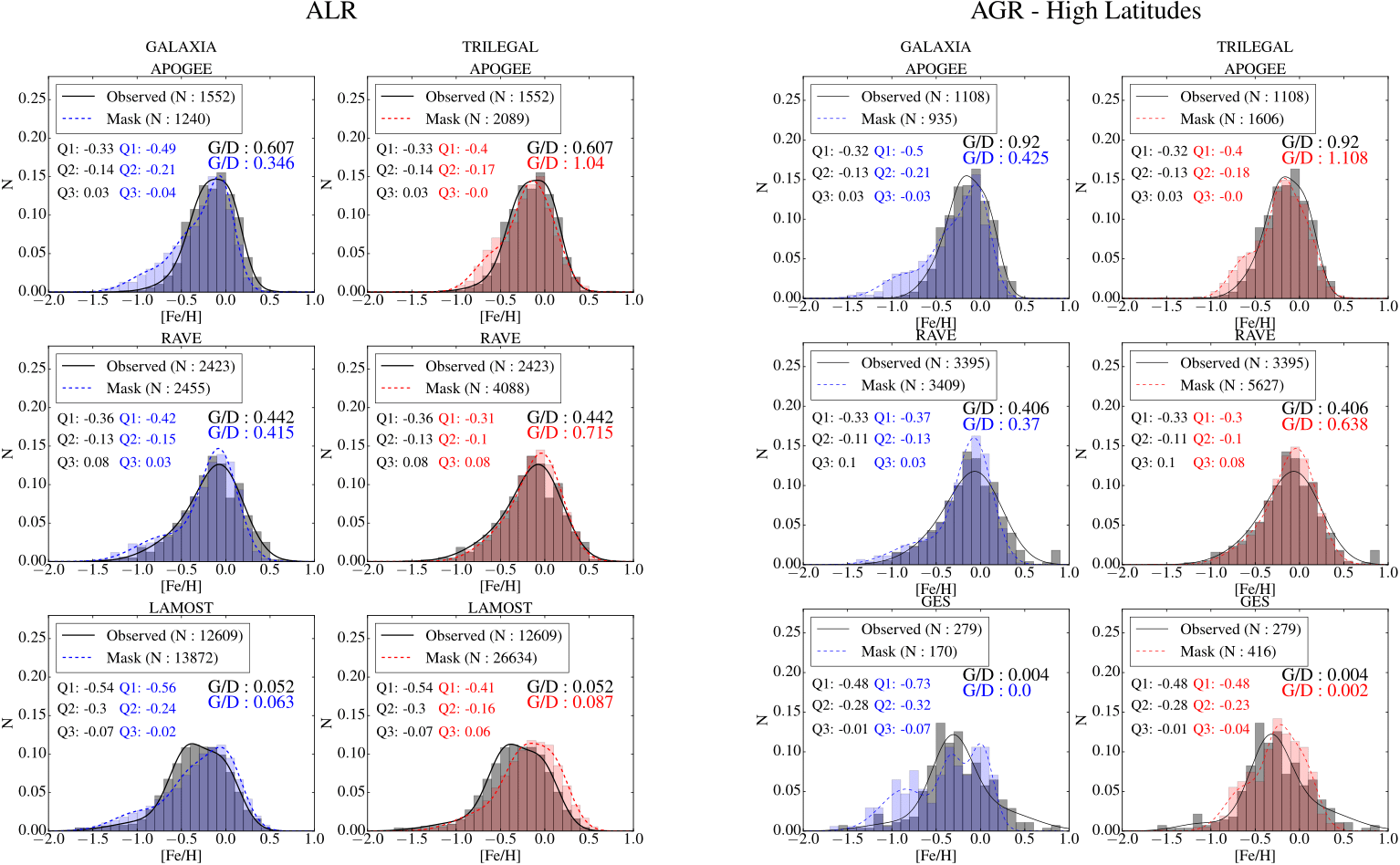}
        \caption{ Mask and observed normalized MDF for ALR (left) and AGR (right) in the R-Z range of 7$\leq$ R $\leq$ 9 kpc and 0$\leq$ $|$Z$|$ $\leq$ 2 kpc. The survey histograms are in black, while GALAXIA and TRILEGAL histograms are in blue and red, respectively. Histograms are normalized by dividing the counts in each 0.1\,dex bin by the total number of sources. The distributions in black line represent the observed MDF, while those in blue and red lines represent the mask MDF for GALAXIA and TRILEGAL, respectively. APOGEE and RAVE distributions are shown in the top and middle rows, respectively, and the LAMOST (left) and GES (right) in the bottom row. Quartile values for both distributions are given in each panel colour-coded according to the  distribution. Indicated is also the giant-to-dwarf ratio for mask and observed samples of each survey for both models.}
  \label{Obs_Mask}
  \end{figure*}

Using the masks in the CMD we force each model to have the same fraction (N$_{observed}$/N$_{photometricsample}$) as the real targeted and observed sample. To understand which model best replicates the observed MDF, we compare the mask and the observed sample MDFs combining all common fields for each survey in ALR and AGR, restricted in the R-Z range of 7$\leq$ R $\leq$ 9 kpc and $|$Z$|$ $\leq$ 2 kpc, as shown in Figure~\ref{Obs_Mask}. Each model uses different stellar libraries and stellar evolutionary tracks,  which could lead to systematic offsets in the abundance scale between the models and observations. However, as discussed in Sections~\ref{sel_MDF} and ~\ref{model_VMG}, we investigate the selection effect within each model.

Except for APOGEE and GES, GALAXIA predicts a slightly higher number of sources, while  TRILEGAL overestimates  the star density  for all four surveys. This difference in the number of predicted sources with respect to the observed sample could be explained by the different assumption in the density normalization of the two models. The fractional percentage of mask sources compared to the observed sources is listed in  Table~\ref{ALR_AGR_fra}, where    the differences between the two models can be clearly seen.
 
The MDFs are binned in 0.1 dex metallicity bins and then normalized by the total number of sources. The fit of the distributions are carried out using a Gaussian mixture model (GMM) (as in \citealt{2017A&A...601A.140R}). A mixture model (M) is a weighted sum of a number of probability distribution functions, while the mixture in a GMM in 2D is defined by a sum of bivariate normal distributions. For a given data structure composed of  certain underlying substructures/features, the parameters ($\mu$, $\sigma$, etc.) that define the best mixture model with a given number of modes is determined using the expectation-maximization algorithm that maximizes the likelihood function of the mixture. Since we do not know a priori the exact number of components in the data, we use the Bayesian Information Criterion (BIC) as a cost function to assess the relative fitting quality between different proposed mixtures and determine the best solution (the one with the lowest BIC value) to the number of Gaussian components that constitutes the distribution.

In order to quantify the differences between the observed and the mask MDFs, we estimate and compare the quartile values for each distribution as in \cite{2017MNRAS.468.3368W}. The quartile values designated as Q1, Q2, Q3 represent the 25${th}$, 50${th}$ and 75${th}$ percentiles of the distribution, respectively, as indicated in each panel of Figure~\ref{Obs_Mask}. We choose 0.1\,dex (considered to be the general metallicity uncertainty) as the threshold for the difference between the respective quartiles of samples below which the distributions are considered to agree. We note the following results by comparing the mask and observed MDFs for each survey (see Figure~\ref{Obs_Mask}) :
\begin{itemize}
\item Both models show a significant metal-poor tail in the mask MDF compared to the APOGEE MDF, which is more prominent in the case of GALAXIA but absent in the observations. This can be seen in the difference in Q1 quartile between GALAXIA and APOGEE distributions, exceeding the adopted threshold of 0.1\,dex.
\item For RAVE, TRILEGAL matches the observed distribution very well for both AGR and ALR. In GALAXIA the mask MDF is distributed as broadly as the RAVE MDF, though there are subtle differences in the source fraction throughout.
\item For LAMOST, the observed MDF has a broad peak that is skewed towards subsolar metallicities, which both the models are unable to replicate. The difference in Q1 and Q2 quartiles exceed the 0.1\,dex threshold between TRILEGAL and LAMOST distributions.
\item  The small sample size in the case of GES makes it hard to decipher the exact shape of the MDFs, especially for GALAXIA. The majority of the 0.1\,dex bins in the GALAXIA MDF have fewer than 15 sources, except for the bins closer to the peak with more than 25 sources. Thus the normalized GALAXIA mask MDF is dominated by noise in the form of multiple peaks, which we are not able to properly fit using GMM. The TRILEGAL mask MDF has better statistics, resulting in significantly less noise in the distribution. This is also evident in the quartile analysis for GALAXIA, with the Q1 quartile difference between GALAXIA and GES distributions exceeding 0.1\,dex, which is not the case with TRILEGAL. 
\end{itemize}

We also estimated the giant-to-dwarf ratio in the model and observed samples in all cases. The giants and dwarfs are separated based on their log $g$ value, i.e. dwarfs : log $g$>3.5 and giants : log $g$<3.5. These ratios show how well each model is able to replicate the observed stellar population. Among the four surveys, the giant fraction is highest in APOGEE followed by RAVE, LAMOST, and GES. We find that both models are unable to represent the ratio we find in the observed sample for APOGEE as GALAXIA gives a lower value, while TRILEGAL overpredicts the ratio by a factor of $\sim$2 in ALR. The prediction by TRILEGAL is close to the observed case for APOGEE in AGR, although still slightly overpredicted. For RAVE, we find the ratio in GALAXIA to be very close to that of the observed case, while TRILEGAL again gives comparatively higher values for both ALR and AGR. The giant-to-dwarf ratios predicted by both GALAXIA and TRILEGAL are quite similar for LAMOST and for GES, though the sample size is quite limited for the GES mask in GALAXIA.

Overall, neither model is able to reproduce both the MDF and the giant-to-dwarf ratio of the APOGEE sample. Both conditions are found to be very consistently satisfied by GALAXIA in the case of RAVE. Even though the shape of the MDFs are slightly different from the observed MDF, we find consistency in the MDF and giant-to-dwarf ratio between GALAXIA and TRILEGAL for LAMOST. In the case of GES, TRILEGAL  reproduces the observed MDF better than GALAXIA, likely because of the lack of targets in the latter model.  

\begin{table}[hbtp!]
\begin{small}
\caption{ Fractional percentage of mask sources compared to the observed sources for GALAXIA and TRILEGAL in ALR and AGR.}
\label{ALR_AGR_fra}
\vspace{0.2cm}
\begin{tabular}{l c c c c c}
 \hline \hline 
ALR & APOGEE ($\%$) & &  RAVE ($\%$) & & LAMOST ($\%$)\\
 
\hline 
\\
 GALAXIA & -20.16 & & +1.32 & & +10.06  \\

 TRILEGAL & +34.41 & & +68.72 & & +111.23 \\
\\ 
  \hline \hline  
AGR & APOGEE ($\%$) & &  RAVE ($\%$) & & GES ($\%$)\\
 
\hline 
\\
 GALAXIA & -15.62 & & +0.41 & & -39.07  \\

 TRILEGAL & +45.04 & & +65.74 & & +49.10  \\
 
\hline 
\end{tabular}
\end{small}
 \end{table}

\subsection{Effects of the selection function in MDF} \label{sel_MDF}  

With the models described above, we are able to study the effect of the selection function on the MDF for the sample from the common fields of each survey in ALR and AGR. We categorize the sources in the mock fields by the limiting magnitude of the respective surveys and restricted in R-Z range as the parent population. This represents the underlying sample from which the selection function in the form of colour and magnitude cuts are applied to create a subset of mask sources. These mask sources in turn represent the observed sources. Thus by comparing the MDF of the underlying sample, hereafter called the magnitude limited sample, with that of the mask sample, we can assess the effect of the selection function, if any, on the underlying MDF for each survey. For ALR and AGR, we restrict both the magnitude limited and mask sample in the R-Z range of 7$\leq$ R $\leq$ 9 kpc and $|$Z$|$ $\leq$ 2 kpc and all fields are combined together. The $|$Z$|$ values for sources in the three low latitude fields in AGR do not exceed  1 kpc for the selected R range. As these low latitude fields have different selection cuts and low numbers of stars, we restrict our analysis only towards high latitude fields.

\begin{figure*}[!htbp]
  \centering
        \includegraphics[width=\textwidth,angle=0]{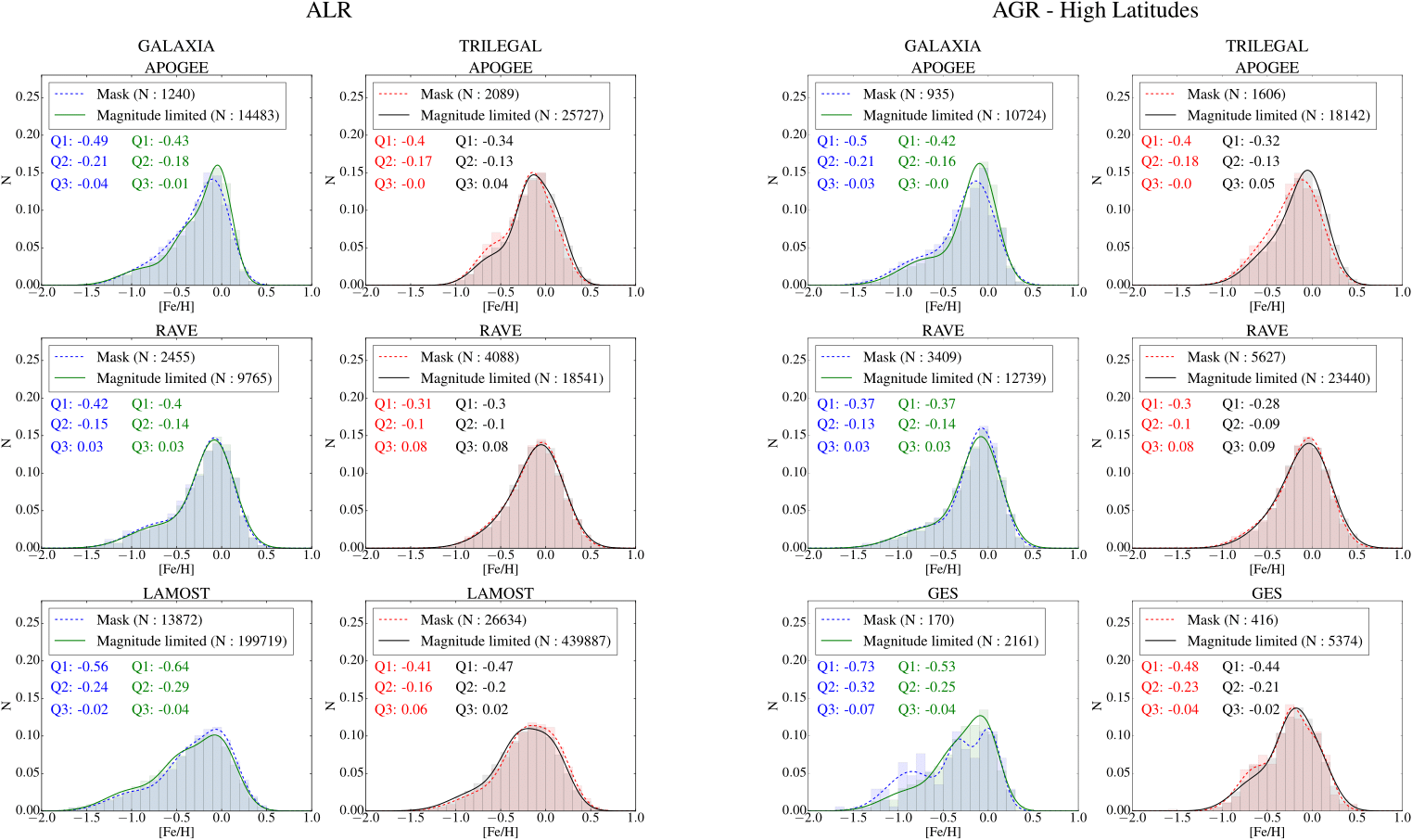}
        
        \caption{ MDFs of magnitude limited and mask sample for the high latitude fields of each survey in ALR (left) and AGR (right). The GALAXIA and TRILEGAL MDFs are shown respectively in the left and right columns of each panel. The histograms are normalized by dividing the counts in each 0.1\,dex bin by the total number of sources (mentioned in each panel). The blue and red lines represent the mask distribution, while the green and black lines for the magnitude limited distribution fitted using GMM for GALAXIA and TRILEGAL respectively. APOGEE and RAVE distributions are shown in the top and middle rows, respectively, and the LAMOST (left) and GES (right) in the bottom row. Quartile values for the two distributions are given in each panel, colour-coded according to the  distribution.}
  \label{GMM_mask_full_ALR}
  \end{figure*}

We compare the magnitude limited MDFs and the effect of the selection function on the MDF for ALR and AGR in Figure~\ref{GMM_mask_full_ALR}. Here again we use the GMM method to fit the multiple number of Gaussians to the MDF. In addition, we use the quartile values to  carry out a quantitive comparison of two distributions. \cite{2017MNRAS.468.3368W} carried out a similar comparison of distributions using the quartile values for RAVE. As mentioned in Section~\ref{models}, we choose 0.1\,dex as the threshold for the difference between the respective quartiles of samples below which the distributions are considered to agree thus implying that the selection function has a minimal effect on the MDF. The quartile values for mask and magnitude limited samples are listed in the respective panels in Figure~\ref{GMM_mask_full_ALR}. We find that all the quartile values estimated for the  mask and the  magnitude limited sample in GALAXIA are more metal-poor than those in TRILEGAL. 
 
\subsubsection*{AGR vs ALR}

As APOGEE and RAVE are the common surveys in ALR and AGR, we check the consistency of the quartile values of their mask and magnitude limited distributions. Since we are using only high latitude fields, the colour and magnitude cuts are consistent for both surveys in ALR and AGR. As a first step, we compare the quartile values of mask and magnitude limited distributions of APOGEE in ALR and AGR for each model separately. The same model-wise quartile comparison is carried out for RAVE in ALR and AGR. We find that model-wise, the individual quartiles differ only by a maximum of $\sim$0.03\,dex for both surveys between ALR and AGR. This ensures that APOGEE and RAVE distributions in ALR and AGR are consistent. 

To quantitively compare the mask and magnitude limited distributions of each model in ALR and AGR, we check for differences in their individual quartiles:
\begin{itemize}

\item For APOGEE, we find the differences in Q1 to be $\sim$0.05-0.08\,dex, for Q2  $\sim$0.03-0.05\,dex, and for  Q3  $\sim$0.03-0.05\,dex. Thus, although we find the quartile differences to vary widely, they are all within the 0.1\,dex difference threshold,  implying no large selection function effect. As mentioned in Section~\ref{APG-selfun}, there are certain halo fields in APOGEE where the giant targets are preselected using the Washington+DDO51 photometry. These photometric bands are not available in the two the models we use,  so we attempted an approximate simulation of this preselection for such fields in our mask sample by using the APOGEE$\_$TARGET1 flag (see Section~\ref{APG-selfun}). We estimated the observed giant-to-dwarf fraction in each small rectangular bins, which we tried to replicate in the models by choosing approximately the same giant fraction. We carried out the comparison of mask and magnitude limited samples with this approximate giant preselection using quartiles. We find there is an overselection of metal-poor stars in the -0.5 to -1.0\,dex range in $\rm [Fe/H]$ in GALAXIA, which is not  evident in TRILEGAL. The Q1 quartiles for mask and magnitude limited samples show differences higher than 0.1\,dex in such fields for GALAXIA, but in Figure~\ref{Obs_Mask}, we find that GALAXIA already (i.e. without any giant preselection) overpredicts the  metal-poor stars in the same $\rm [Fe/H]$ regime compared to the observed APOGEE MDF. It is likely that this selection effect seen in such giant dominated fields is the result of model parameterization, as it already significantly overpredicts contributions from the metal-poor populations.   

\item For RAVE, the individual quartile differences are minimal: $\sim$0.0-0.02\,dex. Thus we find the RAVE MDF to be least affected by the selection effect. \cite{2017MNRAS.468.3368W} have found very similar result for the RAVE DR5 sample, for  separate giant, main-sequence, turn-off samples of stars and for a mixed population sample and at different distance bins from the Galactic mid-plane.

\item Like APOGEE, the individual quartile differences in LAMOST show some variations (Q1: $\sim$0.06-0.08\,dex, Q2: $\sim$0.04-0.05\,dex, and Q3: $\sim$0.03\,dex). However,  as per our criteria, the selection function effect is not prominent. 

\item In the case of GES, we find inconsistency in the quartiles between GALAXIA and TRILEGAL. The quartile differences of the mask and magnitude limited sample in TRILEGAL agrees within the 0.1\,dex threshold (Q1: $\sim$0.04\,dex, Q2: $\sim$0.02\,dex, and Q3: $\sim$0.02\,dex), but the Q1 values in GALAXIA differ by $\sim$0.2\,dex. We find that GALAXIA masks do not have enough sources in the metal-poor regime (170 stars in total), making this difference highly susceptible to Poisson noise. 

\end{itemize}

\section{Vertical metallicity gradients in ALR and AGR} \label{VMG}

We measure the vertical metallicity gradient of our source sample in the solar neighbourhood for each survey in ALR and AGR. We investigate the possible  selection effect on the metallicity  gradient  using both stellar population synthesis models (GALAXIA and TRILEGAL). We further determine the vertical metallicity gradient for each survey independently after accounting for metallicity offsets between them.

\subsection{Effects of selection function in vertical metallicity gradients} \label{model_VMG}

We use stellar population synthesis models as described in Section~\ref{sel_MDF} to  simulate any influence of the selection function on the vertical metallicity gradient.

 Here we estimate and compare the vertical metallicity gradients for the mask and magnitude limited sample of each survey. In  both models, the gradient in metallicity in the vertical direction is not incorporated as an input parameter. Instead, the mean metallicity of different Galaxy components like thin and thick discs, combined with their different scale heights, leads to the vertical metallicity gradient. In TRILEGAL, this is found to be shallow ($\sim$ -0.1\,dex kpc$^{-1}$), while that in GALAXIA is steeper ($\sim$ -0.4\,dex kpc$^{-1}$). This can be attributed to the wide range of the  age--metallicity relation for thin disc in GALAXIA ($\sim$-0.01 to -0.37\,dex \citealt{2011ApJ...730....3S}). Thus  the two models differ largely in their vertical metallicity gradients. We use only the ALR sample to carry out our simulations to ensure sufficient statistics.

\begin{table}[hbtp!]

\caption{Vertical metallicity gradients measured for mask and magnitude sample for GALAXIA and TRILEGAL in ALR. }
\label{Z_MH_models}
\begin{small}
\begin{tabular}{c   c  c  c}
\hline \hline 
 Model&  Survey  & Mask (dex kpc$^{-1}$)  &   Mag limited (dex kpc$^{-1}$)\\

\hline
\\
\multirow{3}{*}{GALAXIA}  & APOGEE &  -0.359$\pm$0.033  & -0.382$\pm$0.025\\
&  LAMOST & -0.396 $\pm$ 0.019 & -0.378 $\pm$ 0.017\\
&  RAVE & -0.438 $\pm$ 0.047 & -0.424 $\pm$ 0.042\\

\hline \hline

\\
\multirow{4}{*}{TRILEGAL}  & APOGEE & -0.085$\pm$0.018 &  -0.054$\pm$0.01 \\
&  LAMOST & -0.037 $\pm$ 0.005 & -0.044 $\pm$ 0.005\\
&  RAVE & -0.121 $\pm$ 0.019 & -0.103 $\pm$ 0.016\\
& GES & -0.075 $\pm$ 0.039 & -0.072 $\pm$ 0.011\\
\hline  
\end{tabular}
\end{small}

 \end{table}

As mentioned in Section~\ref{models}, the mask sample is made by randomly choosing the model sources within each 0.05 mag by 0.3 mag bins in the CMD.  The slopes of the gradients are measured by finding the median metallicity in 0.2 kpc bins in $|$Z$|$ and then using a linear least-squares regression fit to the median values. Unlike in the case of MDF, we find that the source distribution in the $|$Z$|$-$\rm [Fe/H]$ plane is sensitive to the random distribution of stars calculated by the models for the mask sample. We find that the mask gradient varies each time the model is run since the location of mask sample stars at the high Z boundaries keep varying. In order to account for this, we performed ten different mask samples for each survey in ALR. We use the median value of the vertical metallicity gradient and its error estimated from all ten  trials as the final gradient for the mask sample. If there is a major influence of the selection function, we expect to find   different metallicity gradients between the mask and magnitude limited samples. Table~\ref{Z_MH_models} gives the fitted values of the metallicity gradient for ALR using GALAXIA and TRILEGAL. We find that the variation in the gradient estimated between the mask and magnitude limited samples are consistent within the 1-$\sigma$ limit for all three surveys (see Table~\ref{Z_MH_models}). This indicates that the selection function does not have a strong impact on the vertical metallicity gradient. We investigate this further in the following section using the vertical metallicity gradients estimated for the observed sample from each survey.

\subsection{Vertical metallicity gradients for the observed sample} \label{VMGO}

To estimate the vertical metallicity gradient and compare them between the different surveys, we have to ensure that the  metallicities of the different surveys are on the same scale. We applied a small offset to the RAVE and LAMOST metallicities with respect to our reference sample APOGEE using the linear and second-degree polynomial functions that we fitted in Section~\ref{APG-RAVE} and Section~\ref{APG-LAM}, respectively. While estimating the functional relation of metallicity offsets of RAVE and LAMOST with respect to APOGEE, we make sure that the relation holds true seperately for both high and low \textit{S$\rm/$N} samples of RAVE and LAMOST. So we can proceed without any major quality cuts for each survey, ensuring a statistically significant sample for our study. 

\begin{figure*}[!htbp]
  \centering
        \includegraphics[width=\textwidth,angle=0]{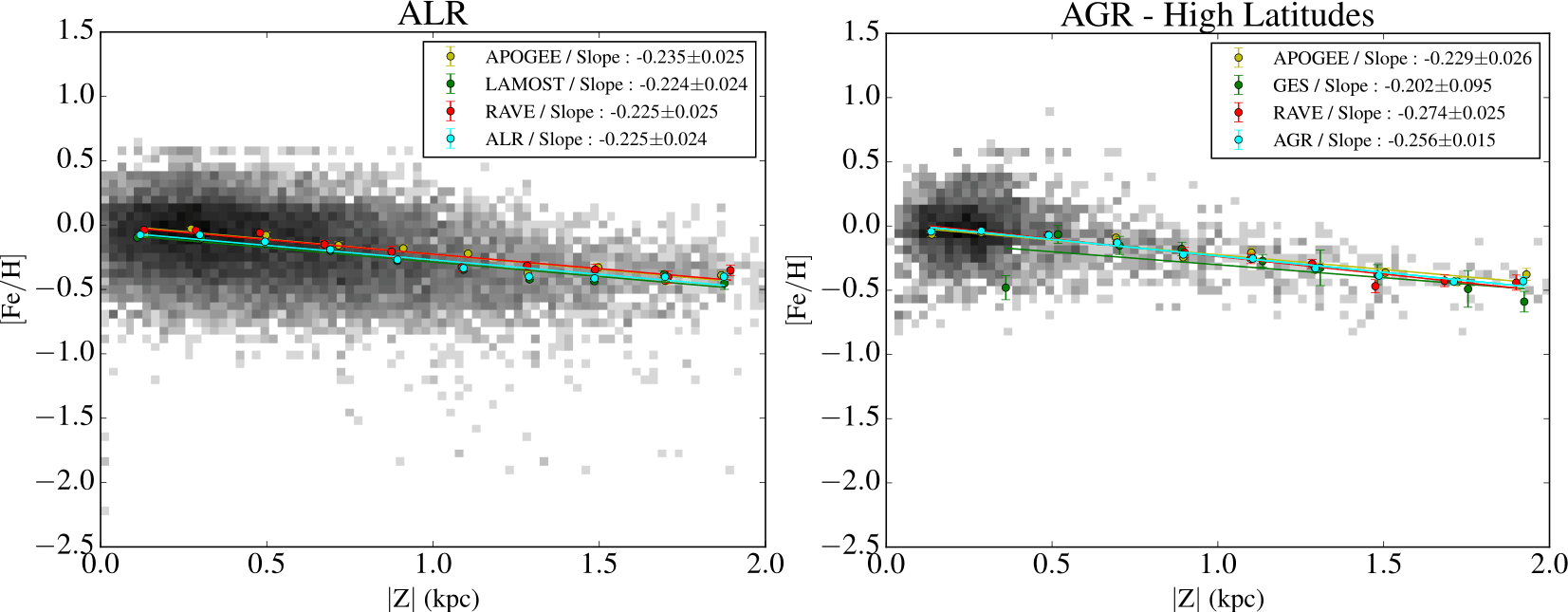}
        \caption{Vertical metallicity gradients calculated for all sources in each survey belonging to ALR (left) and AGR (right). The slope estimated for each survey is also shown in the plots. The gradient for the combined sample of surveys is  shown for ALR and AGR.}
  \label{Z_MH}
  \end{figure*}

Here again we restrict  our study to the high latitude fields. Figure~\ref{Z_MH} shows the vertical metallicity gradients plotted separately for ALR and AGR. We follow the same fitting routine mentioned
in the previous section to estimate the slopes. Table~\ref{Z_MH_tab} lists the slopes of the gradients calculated for each survey in ALR and AGR, along with the slope of the combined sample. We also list the mean vertical metallicity gradient estimated from combined samples of ALR and AGR in the last row of the table.

\begin{table}[hbt!]
\begin{center}
\caption{Vertical metallicity gradients measured for ALR and AGR  high latitude fields.}
\label{Z_MH_tab}
\vspace{0.2cm}
\begin{tabular}{c c | c}

& Survey  & \vspace{0.1cm}$\frac{d\rm [Fe/H]}{dZ}$ (dex kpc$^{-1}$) \\

\hline
\multirow{ 4}{*} & APOGEE & -0.235$\pm$0.025 \\
& LAMOST & -0.224 $\pm$ 0.024\\
& RAVE & -0.225 $\pm$ 0.025\\
\vspace{0.2cm}& \textbf{ALR} & \textbf{-0.225$\pm$ 0.024}\\

\multirow{ 4}{*} & APOGEE & -0.229$\pm$0.026 \\
& GES & -0.202 $\pm$ 0.095\\
& RAVE & -0.274 $\pm$ 0.025\\
& \textbf{AGR} & \textbf{-0.256 $\pm$ 0.015}\\
\hline 
\hline
& &\\
& Mean $\frac{d\rm [Fe/H]}{dZ}$ & \textbf{-0.241$\pm$ 0.028}\\
& &\\
\hline
\end{tabular}
\end{center}
 \end{table}

We measured the vertical gradients of -0.235 $\pm$ 0.025\,dex kpc$^{-1}$ and -0.229 $\pm$ 0.026\,dex kpc$^{-1}$ for APOGEE in ALR and AGR, respectively. \cite{2014AJ....147..116H} measured a slightly steeper slope of -0.31 $\pm$ 0.01\,dex kpc$^{-1}$ for their APOGEE DR10 sample located  within the solar neighbourhood, 7$<$R$<$9 kpc and $|$Z$|$ $<$ 2 kpc. Their sample in the solar neighbourhood is more complete and homogeneous than the volume limited sample we are dealing with. We used the same criterion in \cite{2014AJ....147..116H} to distinguish the $\alpha$-poor and $\alpha$-rich sources in our sample, and measured $\frac{d\rm [M/H]}{dZ}$ = -0.175 $\pm$ 0.045\,dex kpc$^{-1}$ and -0.164 $\pm$ 0.035\,dex kpc$^{-1}$ for the low-$\alpha$ samples in ALR and AGR, respectively. \cite{2014AJ....147..116H} measured  a slightly steeper gradient of -0.21 $\pm$ 0.02\,dex kpc$^{-1}$ for their set
of low-$\alpha$ samples. Meanwhile, the low  number statistics of the high-$\alpha$ sources in our sample prevented us from calculating the same. The vertical metallicity gradient for the DR13 APOGEE sources in the same R-Z range was recalculated and found to be consistent with our slope (Hayden et al. 2017 in prep.). 

The RAVE vertical metallicity gradient calculated for ALR and AGR are also similar, with slopes of -0.225 $\pm$ 0.025\,dex kpc$^{-1}$ and -0.274 $\pm$ 0.025\,dex kpc$^{-1}$, respectively. The steeper slope in AGR could be due to the comparatively low number statistics in the bins at  high $\rm |Z|$ with respect to that in ALR, which makes the slope steeper.  \cite{2014A&A...568A..71B}  used giants stars in the RAVE DR4 sample and measured a shallower slope of $\frac{d\rm [Fe/H]}{dZ}$ = -0.112 $\pm$ 0.007\,dex kpc$^{-1}$ for $\sim$ 10511 stars (RC sample) in the region extending from 7.5 to 8.5 kpc in R and $|$Z$|$ $\leq$ 2 kpc. They have also carried out a study of the gradients seen in the $\alpha$-poor and $\alpha$-rich sample, but we were not able to identify any clear separation   in the [$\alpha$/Fe] vs $\rm [Fe/H]$ plane to define a  $\alpha$-poor and $\alpha$-rich sample.

The low number statistics of GES is a significant issue when analysing metallicity trends. Since we use only the high latitude fields, there are no sources populated  in the first 0.2 kpc bins of $|$Z$|$. We also find a comparatively metal-poor median metallicity value in the 0.2--0.4 kpc $|$Z$|$ bin, which leads to a very high uncertainty of $\sim$ 0.1\,dex kpc$^{-1}$ in the estimated gradient. We calculated $\frac{d\rm [Fe/H]}{dZ}$ of -0.202 $\pm$ 0.095\,dex kpc$^{-1}$, which is still inside  1-$\rm \sigma$ with respect to APOGEE and RAVE.  \cite{2014A&A...572A..33M} measured vertical metallicity gradients of -0.079 $\pm$ 0.013 and -0.046 $\pm$ 0.010\,dex kpc$^{-1}$ for thin and thick-disc FGK stars separately in the solar circle (7$\leq$R$\leq$9 kpc and $|$Z$|$ $\leq$ 3.5 kpc), which are much shallower than our measurements. They use the first internal data release of the GES (GES iDR1) and chemically differentiate the thin and thick discs based on their $\alpha$ abundances, whereas our sample is made using the iDR4 release and we do not make any separation based on the $\alpha$-abundances.

The vertical metallicity gradient measured for LAMOST is steep, -0.224 $\pm$ 0.024\,dex kpc$^{-1}$, but within the 1-$\sigma$ limit with respect to APOGEE and RAVE. Using nearly 70 000 red clump stars covering 7$\leq$R$\leq$14 kpc and $|$Z$|$ $\leq$ 3 kpc from the LAMOST Spectroscopic Survey of the Galactic Anti-Centre (LSS-GAC) survey, \cite{2015RAA....15.1240H} measured the radial and vertical metallicity gradients. Our estimate is steeper than their slopes of -0.146 $\pm$ 0.012\,dex kpc$^{-1}$ and -0.149$\pm$ 0.012\,dex kpc$^{-1}$ measured for the sample in 7$\leq$R$\leq$8 kpc and 8$\leq$R$\leq$9 kpc, respectively. \cite{2015RAA....15.1209X} measured a vertical metallicity gradient that is in the range of $\sim$ -0.2 to -0.3\,dex kpc$^{-1}$ in the R bin of 8 to 9 kpc and $|$Z$|$ $<$ 2 kpc, for a sample of main-sequence turn-off stars from LSS-GAC, which is consistent with our value.

Our mean vertical metallicity gradient from the combined samples of ALR and AGR is -0.241$\pm$0.028\,dex kpc$^{-1}$. Among recent studies, \cite{2014ApJ...791..112S} carried out a detailed study of the vertical metallicity gradient using over 40 000 G-dwarf stars from the SEGUE DR9 catalogue, volume complete in the range of 6.7 to 9.5 kpc in R and 0.27 to 1.62 kpc in $|$Z$|$. Their range in R-Z is very close to the coverage of our samples. They estimated the gradient to be -0.243$^{+0.039}_{-0.053}$\,dex kpc$^{-1}$, which is in good agreement with our derived mean value.

In addition to the results from the surveys used in our analysis, there are other studies calculating the vertical metallicity gradients near the solar circle, \citet{2003AJ....125.1397C}: -0.295 $\pm$ 0.005\,dex kpc$^{-1}$, \citet{2003A&A...398..133B}: -0.29 $\pm$ 0.06\,dex kpc$^{-1}$, \citet{2008A&A...480...91S}: -0.31 $\pm$ 0.03\,dex kpc$^{-1}$ for thin disc clump giants within $|$Z$|$ $\leq$ 1 kpc, \citet{2007NewA...12..605A}: -0.22 $\pm$ 0.03\,dex kpc$^{-1}$ for G-type stars in $|$Z$|$ $\leq$ 3 kpc. All these gradients are very close to the gradients we estimated for each survey. Vertical gradients have been measured for thick discs alone;  \citet{2011AJ....142..184C} have measured a gradient of -0.22 $\pm$ 0.07\,dex kpc$^{-1}$ for RHB stars in 1 $\leq$ $|$Z$|$ $\leq$ 3 kpc, and \citet{ 2012AJ....144..185C} -0.113 $\pm$ 0.010\,dex kpc$^{-1}$ for SEGUE FGK dwarf stars in 1 $\leq$ $|$Z$|$ $\leq$ 3 kpc. \cite{2011A&A...535A.107K} have estimated a vertical metallicity gradient of -0.14 $\pm$ 0.05\,dex kpc$^{-1}$ using roughly 700 stars at  1 $\leq$ $|$Z$|$ $\leq$ 4 kpc  from the solar neighbourhood towards the Galactic coordinates (l$\sim$ 277\,\degr , b$\sim$ 47\,\degr).

While there is a large variation in the vertical metallicity gradient in the literature, we find consistent measurements of the vertical gradient in the four spectroscopic surveys analysed in this paper. This implies that the effect of the selection function on the vertical metallicity gradient is very small, if any at all. 

As seen in Section~\ref{model_VMG} the metallicity gradients for the mask sample in both models (Table~\ref{Z_MH_models}) do not match the observed metallicity gradient (Table~\ref{Z_MH_tab}). However, as we compare  the magnitude limited sample and the mask sample for each model separately to study the influence of the selection effect, we neglect the discrepancy between the observed gradient and both models. Nevertheless, it shows that both models need to be improved in order to reproduce the observed quantities (e.g. metallicity gradient).

Thus, using the models and the observed sample, we find negligible selection function effects on the vertical metallicity gradient. This in turn means that it is indeed possible to merge different spectroscopic surveys to obtain a broader Z range that traces  the vertical metallicity gradient to higher precision, provided they are on the same metallicity scale.

\section{Conclusions} \label{Cncls}

We investigated the effect of the selection function on the MDF and on the vertical metallicity gradient using common fields between APOGEE-LAMOST-RAVE (ALR) and 
APOGEE-GES-RAVE (AGR) around the solar neighbourhood.  Our results can be summarized as follows:

 We compared and discussed stellar parameters of the common sources between APOGEE, RAVE, and LAMOST. In order to bring  the surveys to the same metallicity scale, we  applied corrections for
 $\rm [Fe/H]$ with respect to APOGEE,  which we used as the reference sample. Distances were calculated for all four surveys and we restricted our sample  in R from 7 to 9 kpc and $|$Z$|$ from 0 to 2 kpc allowing the determination of the MDF and vertical metallicity gradient.
 
 We generated mock fields using two commonly used population synthesis models, GALAXIA and TRILEGAL, to investigate the selection effect in MDFs.  We divided the corresponding CMDs into small bins (called masks) which we applied to each  model trying to replicate the observed MDF.
 Based on the comparison of the shape of mask MDFs and  giant-to-dwarf ratio we find that  
\begin{itemize}
\item for APOGEE,  both models have a dominant metal-poor tail absent in the observed MDF. In addition, GALAXIA underestimates the giant-to-dwarf ratio while TRILEGAL overestimates it;
\item GALAXIA traces both the shape of the MDF and the giant-to-dwarf ratio of the RAVE sample better than  TRILEGAL;
\item for LAMOST there is good consistency between
the giant-to-dwarf ratio of mask samples and observed samples, even though the shape of the mask MDF relative to the observed is found to be different in the two models.

\end{itemize}

 To understand the selection function effect in MDF, we compared the mask MDF with the magnitude limited MDF for the survey-replicas of both models using quartiles. We found APOGEE, RAVE, and LAMOST to have negligible selection effects using both models, while GES  suffers from  number statistics that are too low to be conclusive.

 We simulated  the vertical metallicity gradient with the two models, and they both failed to reproduce our observed metallicity gradient;  clearly some improvements in both models are necessary. However, when we compare the vertical metallicity gradients between mask and magnitude limited samples for APOGEE, RAVE, and LAMOST, which are sensitive to the selection function, we did not find any significant difference. In addition, the agreement  found in our observed values of the metallicity gradient  between the four different surveys  again strengthens the argument that the selection effect plays a negligible role when determining the metallicity gradient.

 We scaled the metallicity values in RAVE and LAMOST to that in APOGEE and compared the vertical metallicity gradients for combined fields for each survey in ALR and AGR. The estimated vertical metallicity gradient for each survey is consistent within 1-$\sigma$ indicating the negligible effect of selection function. Finally, we estimated a mean vertical metallicity gradient of -0.241$\pm$0.028\,dex kpc$^{-1}$.

We conclude that in the era of  rising large spectroscopic  surveys, in principle common fields of the surveys could be combined once they are put on the same metallicity scale. This will  increase significantly the statistics without imposing any selection effect when studying the MDF and the metallicity gradient. With the forthcoming Gaia data releases, we plan to extend this study to a much larger volume .

\begin{acknowledgements}
We wish to thank the anonymous referee for the extremely useful comments that greatly improved the quality of this paper. G.N. and M.S. acknowledge the Programme National de Cosmologie et Galaxies (PNCG) of CNRS/INSU, France, for financial support.
\\
Funding for SDSS-III has been provided by the Alfred P. Sloan Foundation, the Participating Institutions, the National Science Foundation, and the U.S. Department of Energy Office of Science. The SDSS-III web site is http://www.sdss3.org/. SDSS-III is managed by the Astrophysical Research Consortium for the Participating Institutions of the SDSS-III Collaboration including the University of Arizona, the Brazilian Participation Group, Brookhaven National Laboratory, Carnegie Mellon University, University of Florida, the French Participation Group, the German Participation Group, Harvard University, the Instituto de Astrofisica de Canarias, the Michigan State/Notre Dame/JINA Participation Group, Johns Hopkins University, Lawrence Berkeley National Laboratory, Max Planck Institute for Astrophysics, Max Planck Institute for Extraterrestrial Physics, New Mexico State University, New York University, The Ohio State University, Pennsylvania State University, University of Portsmouth, Princeton University, the Spanish Participation Group, University of Tokyo, University of Utah, Vanderbilt University, University of Virginia, University of Washington, and Yale University.
\\
Funding for RAVE has been provided by the Australian Astronomical Observatory; the Leibniz-Institut fuer Astrophysik Potsdam (AIP); the Australian National University; the Australian Research Council; the
French National Research Agency; the German Research Foundation (SPP 1177 and SFB 881); the European Research Council (ERC-StG 240271 Galactica); the Istituto Nazionale di Astrofisica at Padova;  Johns Hopkins University; the National Science Foundation of the USA (AST-0908326); the W. M. Keck foundation; the Macquarie University; the Netherlands Research School for Astronomy; the Natural Sciences and Engineering Research Council of Canada; the Slovenian Research Agency; the Swiss National Science Foundation; the Science $\&$ Technology Facilities Council of the UK; Opticon; Strasbourg Observatory; and the Universities of Groningen, Heidelberg, and Sydney.
\\
Based on data products from observations made with ESO Telescopes at the La Silla Paranal Observatory under programme ID 188.B-3002. These data products have been processed by the Cambridge Astronomy Survey Unit (CASU) at the Institute of Astronomy, University of Cambridge, and by the FLAMES/UVES reduction team at INAF/Osservatorio Astrofisico di Arcetri. These data have been obtained from the Gaia-ESO Survey Data Archive, prepared and hosted by the Wide Field Astronomy Unit, Institute for Astronomy, University of Edinburgh, which is funded by the UK Science and Technology Facilities Council. This work was partly supported by the European Union FP7 programme through ERC grant number 320360 and by the Leverhulme Trust through grant RPG-2012-541. We acknowledge the support from INAF and Ministero dell’ Istruzione, dell’ Università’ e della Ricerca (MIUR) in the form of the grant "Premiale VLT 2012". The results presented here benefit from discussions held during the Gaia-ESO workshops and conferences supported by the ESF (European Science Foundation) through the GREAT Research Network Programme.
\\
The Guo Shou Jing Telescope (the Large Sky Area Multi-Object Fiber Spectroscopic Telescope, LAMOST) is a National Major Scientific Project built by the Chinese Academy of Sciences. Funding for the project has been provided by the National Development and Reform Commission. LAMOST is operated and managed by National Astronomical Observatories, Chinese Academy of Sciences.
\end{acknowledgements}

\bibliographystyle{aa}
\bibliography{Selectionfunction}

\end{document}